\documentclass[a4paper,11pt]{article}
\usepackage[toc,page]{appendix}
\usepackage{jheppub}
\usepackage{setspace}
\usepackage{amssymb}
\usepackage{amsmath,amsfonts}
\usepackage{graphicx}
\usepackage{mathrsfs}
\usepackage[vcentermath]{youngtab}
\usepackage{color}
\numberwithin{equation}{section}
\usepackage{bbm}

\newcommand{\be}{\begin{equation}}
\newcommand{\ee}{\end{equation}}
\newcommand{\bea}{\begin{eqnarray}}
\newcommand{\eea}{\end{eqnarray}}

\newcommand{\ket}[1]{\left\lvert #1 \right\rangle}

\usepackage{multirow}

\newcommand{\mfg}{\mathfrak{g}}

\newcommand{\commute}[2]{\left[ #1 \, , \, #2 \right]}
\newcommand{\anticommute}[2]{\left\{ #1 \, , \, #2 \right\}}
\usepackage{array}
\usepackage{graphicx}
\usepackage{longtable}
\usepackage{color}
\usepackage{bbm}
\usepackage{appendix}

\newcommand{\bep}{\begin{picture}}
\newcommand{\eep}{\end{picture}}
\newcounter{YoungHeight}\newcounter{YoungWidth}

\newcounter{Mul1}\newcounter{Mul2}\newcounter{Mul3}\newcounter{Mul4}
\newcounter{A1}\newcounter{A2}
\newcounter{B3}
\newcounter{C3}\newcounter{C4}

\newcounter{T0}\newcounter{T1}

\newcounter{R0}

\newlength{\txtHShift}

\newlength{\txtWidth}

\newcommand{\Add}[3]{\setcounter{#1}{#2}\addtocounter{#1}{#3}}

\newcommand{\Length}[1]{#10}
\newcommand{\YoungScale}{}%\unitlength=0.35mm}

\newcommand{\BlockA}[2]{{\YoungScale\bep(\Length{#1},\Length{#2}){\Add{A1}{#1}{1}\Add{A2}{#2}{1}}%
\multiput(0,0)(10,0){\value{A1}}{\line(0,1){\Length{#2}}}\multiput(0,0)(0,10){\value{A2}}{\line(1,0){\Length{#1}}}%
\setcounter{YoungHeight}{\Length{#2}}\setcounter{YoungWidth}{\Length{#1}}\eep}}

\newcommand{\YoungB}{\BlockA{2}{1}}

\newcommand{\YoungAA}{\BlockA{1}{2}}

\newcommand{\YoungBB}{\BlockA{2}{2}}

\newcommand{\YoungAAAA}{\BlockA{1}{4}}

\usepackage[normalem]{ulem}

\begin{document}

\title{Massless conformal fields, $AdS_{(d+1)}/CFT_d$ higher spin algebras and their deformations}

\author{Sudarshan Fernando$^a$  and  }
\author{Murat G\"unaydin$^{b}$}

\affiliation{$^{a}$  Physical Sciences Department \\
Kutztown University \\
Kutztown, PA 19530, USA}
\affiliation{$^b$ Institute for Gravitation and the Cosmos \\
Physics Department,
Pennsylvania State University \\
University Park, PA 16802, USA}

\emailAdd{fernando@kutztown.edu}
\emailAdd{mgunaydin@psu.edu}
\abstract{We extend our earlier work on the minimal unitary representation of $SO(d,2)$ and   its deformations for $d=4,5$ and $6$  to arbitrary dimensions $d$.  We show that there is a one-to-one correspondence between the minrep of $SO(d,2)$ and its deformations and massless conformal fields in Minkowskian spacetimes in $d$ dimensions. The minrep describes a massless conformal scalar field, and its deformations describe massless conformal fields of higher spin. The generators of Joseph ideal vanish identically as operators for the quasiconformal realization of the minrep, and its enveloping algebra  yields directly the standard bosonic $AdS_{(d+1)}/CFT_d$ higher spin algebra. For deformed minreps  the generators of certain deformations of Joseph ideal vanish as operators and their enveloping  algebras lead to deformations of the standard bosonic higher spin algebra. In odd dimensions there is a unique deformation of the higher spin algebra corresponding to the spinor singleton. In even dimensions one finds infinitely many deformations of the higher spin algebra labelled by the eigenvalues of Casimir operator of the little group $SO(d-2)$ for massless representations.}

\maketitle

\section{Introduction}
\label{sec:intro}

In earlier work we studied the minimal unitary representation (minrep) of $SO(d,2)$ for $d=4,5$ and $6$ and their deformations using the quasiconformal approach. More specifically, in \cite{Fernando:2009fq} we constructed the minrep of $SU(2,2)$  and  its  one-parameter family of deformations which describe all massless conformal fields in four dimensions with the identification of the deformation parameter  as helicity. The minrep of the superalgebras $\mathfrak{su}(2,2|N)$ that extend the conformal algebra in four dimensions were also constructed in  \cite{Fernando:2009fq}. The minrep of $SU(2,2|N)$ also admits an infinite family of deformation which describe massless $N$-extended superconformal multiplets in four dimensions. The minimal unitary supermultiplet of the $N=4$ superconformal algebra $PSU(2,2|4)$ is simply the $N=4$ Yang-Mills supermultiplet.

These results were later extended to the construction of the minimal unitary representation of the six-dimensional conformal group $SO(6,2) \approx SO^*(8)$ and its deformations and their supersymmetric extensions \cite{Fernando:2010dp,Fernando:2010ia}.  There exists a discrete infinite family of deformations of the minrep of $SO(6,2)$ , labelled by the eigenvalues of an $SU(2)$ subgroup of the little group $SO(4)$ of massless particles in $6d$. This infinite family of deformations of the minrep of $SO^*(8)$ describe massless conformal fields in six dimensions. The minimal unitary supermultiplets of $6d$  conformal superalgebras $\mathfrak{osp}(8^*|2N)$ also admit a discrete infinite family of deformations which  describe massless conformal supermultiplets in six space-time dimensions.

In more recent work we constructed  the minimal unitary representation of $SO(5,2)$ using quasiconformal methods and showed that it admits a single deformation \cite{Fernando:2014pya}. The minrep of $SO(5,2)$ and its deformation are the analogs of scalar and spinor singletons of the three-dimensional conformal group $SO(3,2)$, which is isomorphic to $Sp(4,\mathbb{R})$, and hence we referred to them as such. The Lie algebra of $SO(5,2)$ admits a unique supersymmetric extension, namely the exceptional Lie superalgebra $\mathfrak{f}(4)$ with the even subalgebra $\mathfrak{so}(5,2) \oplus \mathfrak{su}(2)$. The minimal unitary supermultiplet of $F(4)$ consists of the spinor singleton together with two copies of the scalar singleton \cite{Fernando:2014pya}.

The minrep of $SU(2,2|N)$ and its deformations turn out to be isomorphic to the doubleton supermultiplets for integer and half-integer values of helicity \cite{Fernando:2009fq} that were constructed and studied using twistorial oscillators some time ago \cite{Gunaydin:1984fk,Gunaydin:1998jc,Gunaydin:1998sw}. The minimal unitary supermultiplet corresponds to the unique CPT self-conjugate doubleton supermultiplet.  The twistorial oscillator method was used to obtain, for the first time, the Kaluza-Klein spectrum of IIB supergravity over $AdS_5 \times S^5$  simply by the tensoring of the CPT self-conjugate doubleton supermultiplet of $PSU(2,2|4)$ \cite{Gunaydin:1984fk}. The CPT self-conjugate doubleton of $PSU(2,2|4)$ itself decouples from the Kaluza-Klein spectrum as gauge modes. Its Poincar\'{e} limit is singular and its field theory lives on the boundary of $AdS_5$ as conformally invariant  $N=4$ super Yang-Mills theory as was first pointed out in \cite{Gunaydin:1984fk}.

The minrep of $OSp(8^*|2N)$ and its deformations turn out to be isomorphic to the doubleton supermultiplets constructed using the twistorial oscillators  \cite{Gunaydin:1984wc,Gunaydin:1999ci,Fernando:2001ak}. The Kaluza-Klein spectrum of eleven-dimensional supergravity over $AdS_7 \times S^4$ was similarly obtained by the tensoring of the CPT self-conjugate doubleton supermultiplet of $OSp(8^*|4)$ \cite{Gunaydin:1984wc}. The CPT self-conjugate doubleton supermultiplet does not have a Poincar\'{e} limit and its field theory lives on the boundary of $AdS_7$ as a conformally invariant field theory as was first pointed out in \cite{Gunaydin:1984wc}.

The physical importance of the four-dimensional $N=4$ Yang-Mills supermultiplet and of the six-dimensional $(2,0)$ conformal supermultiplet became abundantly clear after the fundamental paper of Maldacena \cite{Maldacena:1997re} who proposed the duality between IIB superstring theory over $AdS_5\times S^5$ and $SU(\mathcal{N})$ $N=4$ Yang-Mills theory and M-theory over $AdS_7 \times S^4$ and the interacting six-dimensional conformal theory of $(2,0)$ multiplets of Witten \cite{Witten:1995zh}. From a mathematical point of view their importance lies in the fact that
they correspond to the minimal unitary supermultiplets of the respective symmetry superalgebras of $PSU(2,2|4)$ and $OSp(8^*|4)$, respectively.

The minimal unitary realizations  of $SU(2,2|N)$ and of $OSp(8^*|2N)$ and their deformations obtained via quasiconformal methods \cite{Fernando:2009fq,Fernando:2010dp,Fernando:2010ia} were reformulated as bilinears of deformed twistorial oscillators  which transform nonlinearly under the Lorentz group in \cite{Govil:2013uta,Govil:2014uwa}.  Furthermore it was  shown that the enveloping algebras of the minimal unitary realizations of $SU(2,2)$ and $SO^*(8)$ thus obtained lead directly to  the higher spin algebras in $AdS_5$ and $AdS_7$.

 As was first pointed out in \cite{Gunaydin:1989um}, the higher spin algebra in $AdS_4$ as studied by Fradkin, Vasiliev and collaborators \cite{Fradkin:1986qy,Konshtein:1988yg} is simply the enveloping algebra of the scalar singleton of $SO(3,2)$. Later Vasiliev showed that the higher spin algebra in $AdS_{(d+1)}$, for general $d$, is given by the enveloping algebra of $SO(d,2)$ quotiented by the ideal that annihilates  the scalar ``singleton'' representation \cite{Vasiliev:1999ba}. That this ideal is simply the Joseph ideal of the minimal unitary representation of $SO(d,2)$ was established by Eastwood \cite{Eastwood:2002su}.
Undeformed higher spin algebras in  $d$ dimensions were further investigated by Vasiliev in \cite{Vasiliev:2004cm}.

For  symplectic groups $Sp(2n,\mathbb{R})$  the minrep is simply the scalar singleton and it admits a single deformation which is the spinor singleton. Quasiconformal realizations of the singletons of $Sp(2n,\mathbb{R})$ coincide with their realizations as bilinears of  oscillators transforming covariantly under the maximal compact subgroup $U(n)$\cite{Gunaydin:2006vz}. The Joseph ideal vanishes identically for the singletons \cite{Govil:2013uta}. As a consequence  the enveloping algebra of  the $AdS_4$ group $SO(3,2)\equiv Sp(4,\mathbb{R})$ realized as bilinears of covariant twistorial oscillators leads directly to the Fradkin-Vasiliev higher spin algebra as was first pointed out in \cite{ Gunaydin:1989um}. On the other hand, for the doubletonic realizations of the minreps of $SO(4,2)$ and  $SO(6,2)$ as bilinears of  twistorial oscillators transforming covariantly under the respective Lorentz groups, the Joseph ideal does not vanish identically as  operators. However it was shown in \cite{Govil:2013uta,Govil:2014uwa} that for the minreps of $SO(4,2)$ and  $SO(6,2)$ obtained by quasiconformal methods, the Joseph ideals vanish identically as operators. Therefore their enveloping algebras provide unitary realizations of the bosonic higher spin algebras in $AdS_5$ and $AdS_7$, respectively. The enveloping algebras of the deformed minreps of $SU(2,2)$ and of $SO^*(8)$ and their supersymmetric extensions lead to   infinite families of higher spin algebras and superalgebras in $AdS_5$ and  $AdS_7$, respectively \cite{Govil:2013uta,Govil:2014uwa}.

Similarly the Joseph ideal for the minrep of $SO(5,2)$ obtained by the quasiconformal methods vanishes identically as operators and its enveloping algebra yields the bosonic higher spin algebra in $AdS_6$ \cite{Fernando:2014pya}. There is a unique deformed higher spin algebra given by the enveloping algebra of the spinor singleton and the enveloping algebra of the minrep of the Lie superalgebra $\mathfrak{f}(4)$ yields the unique higher spin superalgebra in $AdS_6$.

In this paper we extend our previous work to all $d$-dimensional conformal groups $SO(d,2)$ and construct their minimal unitary representations and their deformations. We shall first review the quasiconformal approach to the construction of the  minreps of $SO(d,2)$\cite{Gunaydin:2006vz} which describe massless conformal fields.   This is followed by the  study of possible deformations of these minreps. In  odd $d$ dimensions, there exists  \emph{a single deformation} that describes  a conformally massless spinor field. 
 In even $d$ dimensions, we find \emph{infinitely many deformations} of the minrep, corresponding to conformally massless fields that transform nontrivially under the little group $SO(d-2)$.  The generators of the Joseph ideal vanish identically as operators for the quasiconformal realization of the minrep of $SO(d,2)$ in all dimensions. Therefore the enveloping algebra of the minrep of $SO(d,2)$ leads directly to the standard bosonic higher spin algebra  in $AdS_{(d+1)}$. For the deformed minreps  certain deformations of the Joseph ideal  vanish identically as operators and their enveloping algebras lead to deformations of the standard bosonic higher spin algebra. The deformations of higher spin algebras  correpond to the  quotients of the  the enveloping algebra of $SO(d,2)$ by the deformations of their Joseph ideals.

The plan of the paper is as follows. Following \cite{Gunaydin:2006vz}, we construct the geometric realization of $SO(d,2)$ as a quasiconformal group in section \ref{sec:geomSO(d,2)}. Then, by quantizing the geometric quasiconformal action, we obtain the minimal unitary representation (minrep) of $SO(d,2)$ in section \ref{sec:minrepSO(d,2)}. In that section we also show that, according to Joseph's theorem \cite{Joseph:1974}, there exists a two-parameter family of degree-two polynomials of $\mathfrak{so}(d,2)$ generators which reduces to a $c$-number. Then in section \ref{sec:SO(d,2)nc3G}, we give the \emph{noncompact} 3-grading of $\mathfrak{so}(d,2)$ with respect to the subalgebra $\mathfrak{so}(d-1,1) \oplus \mathfrak{so}(1,1)$ and show that the Poincar\'{e} mass operator in $d$ dimensions vanishes identically for the minrep of $SO(d,2)$. Similarly in section \ref{sec:SO(d,2)c3G} we give the \emph{compact} 3-grading of $\mathfrak{so}(d,2)$ with respect to the subalgebra $\mathfrak{so}(d) \oplus \mathfrak{so}(2)$ and obtain the Poincar\'{e}-masslessness condition in the compact basis. Then we discuss the properties of a distinguished $SU(1,1)$ subgroup of $SO(d,2)$, which is generated by the longest root vector and realized by singular (isotonic) oscillators, in section \ref{sec:SU(1,1)_K} and present the K-type decomposition of the minrep of $SO(d,2)$ in section \ref{sec:undeformedminrep}. We also show that the minrep of $SO(d,2)$ corresponds to a massless conformal scalar field in $d$ dimensions. Then in section \ref{sec:deformedSO(d,2)} we introduce the deformations of the minrep of $SO(d,2)$. In particular, we present a constraint that needs to be satisfied by the generators that introduce deformations. Then we construct the K-type decomposition of the \emph{unique} deformed minrep for odd $d$ in section \ref{subsec:OddDeformations} and the K-type decomposition of the infinitely many deformed minreps for even $d$ in section \ref{subsec:EvenDeformations}. In section \ref{sec:bosonicHS}, after presenting the $SO(d,2)$-covariant generators in terms of $SO(d) \times SO(2)$-covariant generators and $SO(d-1) \times SO(1,1)$-covariant generators, we show that the Joseph ideal vanishes identically as an operator for the minrep of $SO(d,2)$ and therefore its universal enveloping algebra yields directly the bosonic $AdS_{(d+1)}/CFT_d$ higher spin algebra. We also show that for the deformations of the minrep, a certain deformation of the Joseph ideal vanishes identically and therefore its enveloping algebra yields the deformed $AdS_{(d+1)}/CFT_d$ higher spin algebras. Finally we have some concluding comments in section \ref{sec:conclusions}. Appendices \ref{app:S_ijForOdd} and \ref{app:S_ijForEven} outline how to realize the ``spin'' operators that extend the little group $SO(d-2)$, which allows us to obtain the deformations of the minrep.

%%%%%%%%%%%%%%%%%%%%%%%%%%%%%%%%%%%%%%%%%%%%%%%%%%
%%%%% Geometric realization of SO(d,2) %%%%%%%%%%%%%%%%%%%%%%%%%%%%
%%%%%%%%%%%%%%%%%%%%%%%%%%%%%%%%%%%%%%%%%%%%%%%%%%

\section{Geometric realization of $SO(d,2)$ as a quasiconformal group}
\label{sec:geomSO(d,2)}

We begin by reviewing the geometric quasiconformal realization of $SO(d,2)$ that was presented in \cite{Gunaydin:2005zz}. The Lie algebra $\mathfrak{so}(d,2)$ admits a 5-grading with respect to its subalgebra $\mathfrak{so}(1,1) \oplus \mathfrak{so}(d-2) \oplus \mathfrak{sp}(2,\mathbb{R})$ as follows:
\begin{equation}
\mathfrak{so}(d,2)
= \mathbf{1}^{(-2)} \oplus
  \left( \mathbf{d-2} , \mathbf{2} \right)^{(-1)} \oplus
  \left[ \,
   \Delta \oplus
   \mathfrak{so}(d-2) \oplus
   \mathfrak{su}(1,1)
  \, \right] \oplus
  \left( \mathbf{d-2} , \mathbf{2} \right)^{(+1)} \oplus
  \mathbf{1}^{(+2)}
\end{equation}
where the five grading is determined by  the $\mathfrak{so}(1,1)$ generator $\Delta$. The non-linear quasiconformal group action of $SO(d,2)$  is generated by nonlinear differential operators acting on a $\left( 2 d -3 \right)$-dimensional space $\mathcal{T}$  corresponding to the Heisenberg subalgebra generated by the elements of the subspace $\left[ \mathfrak{g}^{(-2)} \oplus \mathfrak{g}^{(-1)} \right]$. We shall denote the coordinates of the space $\mathcal{T}$ as $\mathcal{X}= \left( X^{i,a} , x \right)$, where $X^{i,a}$ transform in the $(d-2,2)$ representation of $\mathfrak{so}(d-2) \oplus \mathfrak{su}(1,1)$ subalgebra, with $i=1,\dots,d-2$ and $a=1,2$, and $x$ is a singlet coordinate.

There exists a quartic polynomial of the coordinates $X^{i,a}$
\begin{equation}
\mathcal{I}_4 (X)
= \eta_{ij} \eta_{kl} \epsilon_{ac} \epsilon_{bd}
  X^{i,a} X^{j,b} X^{k,c} X^{l,d}
\end{equation}
where $i,j,k,l=1,\dots,d-2$ and $a,b,c,d=1,2$, which is an invariant of $SO(d-2) \times SU(1,1)$ subgroup. In the above expression, $\epsilon_{ab}$ is the symplectic invariant tensor of $SU(1,1)$ and $\eta_{ij}$ is the invariant metric of $SO(d-2)$ in the fundamental representation, which we choose as $\eta_{ij} = -\delta_{ij}$ following the general conventions of \cite{Gunaydin:2005zz}.

We shall label the generators that belong to various grade subspaces as follows:
\begin{equation}
\mathfrak{so}(d,2)
= K_{-} \oplus
  U_{i,a} \oplus
  \left[ \Delta \oplus \mathcal{L}_{ij} \oplus \mathcal{M}_{ab} \right] \oplus
  \widetilde{U}_{i,a} \oplus
  K_{+}
\end{equation}
where $\mathcal{L}_{ij}$ and $\mathcal{M}_{ab}$ are the generators of $SO(d-2)$ and $SU(1,1)$ subgroups, respectively. In the nonlinear quasiconformal action of $SO(d,2)$, these generators take on the form
\begin{equation} \label{QCGaction}
\begin{split}
K_+
&= \frac{1}{2} \left( 2 x^2 - \mathcal{I}_4 \right) \frac{\partial}{\partial x}
   - \frac{1}{4} \frac{\partial \mathcal{I}_4}{\partial X^{i,a}}
     \eta^{ij} \epsilon^{ab} \frac{\partial}{\partial X^{j,b}}
   + x \, X^{i,a} \frac{\partial}{\partial X^{i,a}}
\\
U_{i,a}
&= \frac{\partial}{\partial X^{i,a}}
   - \eta_{ij} \epsilon_{ab} \, X^{j,b} \frac{\partial}{\partial x}
\\
\mathcal{L}_{ij}
&= \eta_{ik} X^{k,a} \frac{\partial}{\partial X^{j,a}}
   - \eta_{jk} X^{k,a} \frac{\partial}{\partial X^{i,a}}
\\
\mathcal{M}_{ab}
&= \epsilon_{ac} X^{i,c} \frac{\partial}{\partial X^{i,b}}
   + \epsilon_{bc} X^{i,c} \frac{\partial}{\partial X^{i,a}}
\\
K_-
&= \frac{\partial}{\partial x}
\qquad , \qquad
\Delta
= 2 \, x \frac{\partial}{\partial x}
  + X^{i,a} \frac{\partial}{\partial X^{i,a}}
\qquad , \qquad
\widetilde{U}_{i,a}
= \commute{U_{i,a}}{K_+}
\end{split}
\end{equation}
where $\epsilon^{ab}$ is the inverse symplectic tensor, such that $\epsilon^{ab} \epsilon_{bc} = {\delta^a}_c$. The explicit form of the grade +1 generators $\widetilde{U}_{i,a}$ can be obtained by substituting the expression for the quartic invariant:
\begin{equation}
\begin{split}
\widetilde{U}_{i,a}
&= \eta_{ij} \epsilon_{ad}
   \left(
    \eta_{kl} \epsilon_{bc} X^{j,b} X^{k,c} X^{l,d}
    - x \, X^{j,d}
   \right)
   \frac{\partial}{\partial x}
   +  x \frac{\partial}{\partial X^{i,a}}
\\
& \quad
   - \eta_{ij} \epsilon_{ab} \, X^{j,b} X^{l,c}
     \frac{\partial}{\partial X^{l,c}}
   - \epsilon_{ad} \eta_{kl} \, X^{l,d} X^{k,c}
     \frac{\partial}{\partial X^{i,c}}
\\
& \quad
   + \epsilon_{ad} \eta_{ij} \, X^{l,d} X^{j,b}
     \frac{\partial}{\partial X^{l,b}}
   + \eta_{ij} \epsilon_{bc} X^{j,b} X^{l,c}
     \frac{\partial}{\partial X^{l,a}}
\end{split}
\end{equation}

The above $SO(d,2)$ generators satisfy the following commutation relations:
\begin{subequations}
\label{eq:sod2algebra}
\begin{equation}
\begin{split}
\commute{\mathcal{L}_{ij}}{\mathcal{L}_{kl}}
&= \eta_{jk} \mathcal{L}_{il} - \eta_{ik} \mathcal{L}_{jl} - \eta_{jl} \mathcal{L}_{ik} + \eta_{il} \mathcal{L}_{jk}
\\
\commute{\mathcal{M}_{ab}}{\mathcal{M}_{cd}}
&= \epsilon_{cb} \mathcal{M}_{ad} + \epsilon_{ca} \mathcal{M}_{bd}
   + \epsilon_{db} \mathcal{M}_{ac} + \epsilon_{da} \mathcal{M}_{bc}
\end{split}
\end{equation}
\begin{equation}
\begin{split}
\commute{\Delta}{K_\pm}
&= \pm 2 \, K_\pm
\qquad \qquad \qquad
\commute{K_-}{K_+}
= \Delta
\\
\commute{\Delta}{U_{i,a}}
&= - U_{i,a}
\qquad \qquad \qquad \quad
\commute{\Delta}{\widetilde{U}_{i,a}}
= \widetilde{U}_{i,a}
\\
\commute{U_{i,a}}{K_+}
&= \widetilde{U}_{i,a}
\qquad \qquad \qquad \quad
\commute{\widetilde{U}_{i,a}}{K_-}
= - U_{i,a}
\\
\commute{U_{i,a}}{U_{j,b}}
&= 2 \, \eta_{ij} \epsilon_{ab} \, K_-
\qquad \qquad
\commute{\widetilde{U}_{i,a}}{\widetilde{U}_{j,b}}
= 2 \, \eta_{ij} \epsilon_{ab} \, K_+
\end{split}
\end{equation}
\begin{equation}
\begin{split}
\commute{\mathcal{L}_{ij}}{U_{k,a}}
&= \eta_{jk} U_{i,a} - \eta_{ik} U_{j,a}
\qquad \qquad
\commute{\mathcal{L}_{ij}}{\widetilde{U}_{k,a}}
= \eta_{jk} \widetilde{U}_{i,a} - \eta_{ik} \widetilde{U}_{j,a}
\\
\commute{\mathcal{M}_{ab}}{U_{i,c}}
&= \epsilon_{cb} U_{i,a} + \epsilon_{ca} U_{i,b}
\qquad \qquad
\commute{\mathcal{M}_{ab}}{\widetilde{U}_{i,c}}
= \epsilon_{cb} \widetilde{U}_{i,a} + \epsilon_{ca} \widetilde{U}_{i,b}
\end{split}
\end{equation}
\begin{equation}
\commute{U_{i,a}}{\widetilde{U}_{j,b}}
= \eta_{ij} \epsilon_{ab} \, \Delta
  - 2 \, \epsilon_{ab} \mathcal{L}_{ij}
  - \eta_{ij} \mathcal{M}_{ab}
\end{equation}
\end{subequations}
The quartic norm (length) of a vector $\mathcal{X}= \left( X^{i,a} , x \right) \in \mathcal{T}$ is defined as
\begin{equation}
\ell \left( \mathcal{X} \right)
= \mathcal{I}_4 \left( X \right) + 2 \, x^2 \,.
\end{equation}
In order to see the geometric picture behind the above nonlinear realization, a quartic distance function between any two points $\mathcal{X}$ and $\mathcal{Y}$ in the $(2d-3)$-dimensional space $\mathcal{T}$ can be defined as \cite{Gunaydin:2001bt,Gunaydin:2006vz}
\begin{equation}
d \left( \mathcal{X} , \mathcal{Y} \right)
= \ell \left( \delta \left( \mathcal{X} , \mathcal{Y} \right) \right)
\end{equation}
where the ``symplectic'' difference $\delta \left( \mathcal{X} , \mathcal{Y} \right)$ is given by
\begin{equation}
\delta \left( \mathcal{X} , \mathcal{Y} \right)
= \left( X^{i,a} - Y^{i,a} \,,\, x - y - \eta_{ij} \epsilon_{ab} \, X^{i,a} Y^{j,b} \right) \,.
\end{equation}
The quasiconformal action of $SO(d,2)$ leaves the lightlike separations between any two points with respect to the quartic distance function invariant, which implies that $SO(d,2)$ behaves as the invariance group of a ``light-cone'' with respect to a quartic distance function in a $\left( 2d - 3 \right)$-dimensional space.

%%%%%%%%%%%%%%%%%%%%%%%%%%%%%%%%%%%%%%%%%%%%%%%%%%
%%%%% Minrep of SO(d,2) %%%%%%%%%%%%%%%%%%%%%%%%%%%%%%%%%%%
%%%%%%%%%%%%%%%%%%%%%%%%%%%%%%%%%%%%%%%%%%%%%%%%%%

\section{Minimal unitary representation of $SO(d,2)$ from its quasiconformal realization}
\label{sec:minrepSO(d,2)}

The minimal unitary realization of a Lie algebra can be easily obtained by quantizing its geometric quasiconformal realization \cite{Gunaydin:2001bt,Gunaydin:2004md,Gunaydin:2005zz,Gunaydin:2006vz}. To achieve this, in the case of $SO(d,2)$, we split the $2(d-2)$ variables $X^{i,a}$ introduced in section \ref{sec:geomSO(d,2)} into $(d-2)$ coordinates $X^i$ and $(d-2)$ conjugate momenta $P_i$ as
\begin{equation}
X^i = X^{i,1}
\qquad \qquad \qquad
P_i = \eta_{ij} \, X^{j,2}
\end{equation}
and introduce a  momentum $p$ conjugate to the singlet coordinate $x$ as well. Treated as quantum mechanical operators, these coordinates and momenta satisfy the canonical commutation relations
\begin{equation}
\commute{X^i}{P_j} = i \, \delta^i_j
\qquad \qquad \qquad
\commute{x}{p} = i \,.
\end{equation}

However, for the rest of this paper, instead of using $X^i$ and $P_i$ we shall work with  bosonic oscillator annihilation operators $a_i$ and creation operators $a_i^\dag$, defined as
\begin{equation}
a_i = \frac{1}{\sqrt{2}} \left( X^i + i \, P_i \right)
\qquad \qquad \qquad
a_i^\dag = \frac{1}{\sqrt{2}} \left( X^i - i \, P_i \right)
\end{equation}
which satisfy the commutation relations
\begin{equation}
\commute{a_i}{a_j^\dag} = \delta_{ij}
\qquad \qquad \qquad
\commute{a_i}{a_j} = \commute{a_i^\dag}{a_j^\dag} = 0 \,.
\end{equation}

The generators of the minimal unitary realization of $\mathfrak{so}(d,2)$ has a  5-graded decomposition
\begin{equation}
\mathfrak{so}(d,2)
= \mathfrak{g}^{(-2)} \oplus \mathfrak{g}^{(-1)} \oplus
  \left[ \,
   \Delta \oplus \mathfrak{so}(d-2) \oplus \mathfrak{su}(1,1)
  \, \right] \oplus
  \mathfrak{g}^{(+1)} \oplus \mathfrak{g}^{+2)}
\end{equation}
with respect to the $SO(1,1)$ generator
\begin{equation}
\Delta = \frac{1}{2} \left( x p + p x \right) \,.
\label{delta}
\end{equation}

The generators of the subalgebra $\mathfrak{su}(1,1) \subset \mathfrak{g}^{(0)}$ are realized as bilinears of the $a$-type bosonic oscillators:
\begin{equation}
M_+ = \frac{1}{2} a_i^\dag a_i^\dag
\qquad
M_- = \frac{1}{2} a_i a_i
\qquad
M_0 = \frac{1}{4} \left( a_i^\dag a_i + a_i a_i^\dag \right)
\label{SU(1,1)M}
\end{equation}
and satisfy
\begin{equation}
\commute{M_-}{M_+} = 2 \, M_0
\qquad \qquad
\commute{M_0}{M_\pm} = \pm \, M_\pm \,.
\end{equation}
We denote this subalgebra as $\mathfrak{su}(1,1)_M$ and its quadratic Casimir as $\mathcal{M}^2$:
\begin{equation}
\mathcal{C}_2 \left[ \mathfrak{su}(1,1)_M \right]
= \mathcal{M}^2
= {M_0}^2 - \frac{1}{2} \left( M_+ M_- + M_- M_+ \right)
\end{equation}

The subalgebra $\mathfrak{so}(d-2) \subset \mathfrak{g}^{(0)}$ corresponding to the little group of massless particles in $d$ dimensions, denoted as $\mathfrak{so}(d-2)_L$, is also realized as bilinears of the $a$-type bosonic oscillators\footnote{ We should note that $L_{ij}$ are Hermitian generators while $\mathcal{L}_{ij}$ in the geometric quasiconformal realization in equation (\ref{QCGaction}) were chosen to be anti-Hermitian.}:
\begin{equation}
L_{ij} = i \left( a_i^\dag a_j - a_j^\dag a_i \right)
\end{equation}
and satisfy the commutation relations
\begin{equation}
\commute{L_{ij}}{L_{kl}}
= i \left( \delta_{jk} L_{il} - \delta_{ik} L_{jl} - \delta_{jl} L_{ik} + \delta_{il} L_{jk} \right) \,.
\end{equation}
The quadratic Casimir $\mathcal{L}^2$ of $\mathfrak{so}(d-2)_L$   given by
\begin{equation}
\mathcal{C}_2 \left[ \mathfrak{so}(d-2)_L \right]
= \mathcal{L}^2
= L_{ij} L_{ij}
\end{equation}
is related to that of $\mathfrak{su}(1,1)_M$ as
\begin{equation}
\mathcal{L}^2 = 8 \, \mathcal{M}^2 - \frac{1}{2} \left( d - 2 \right) \left( d - 6 \right) \,.
\end{equation}

The single generator in $\mathfrak{g}^{(-2)}$ is defined as
\begin{equation}
K_- = \frac{1}{2} x^2 \,.
\label{K-}
\end{equation}
The ``quantized generators''  $(U_i , U_i^\dagger)$ in grade $-1$ subspace are realized as bilinears of $x$ and the above $a$-type bosonic oscillators as
\begin{equation}
U_i = x \, a_i
\qquad \qquad \qquad
U_i^\dag = x \, a_i^\dag \,.
\label{Grade-1Bosonic}
\end{equation}
They close into $K_-$ under commutation and  form a Heisenberg subalgebra
\begin{equation}
\commute{U_i}{U_j^\dag} = 2 \, \delta_{ij} \, K_-
\qquad \qquad
\commute{U_i}{U_j} = \commute{U_i^\dag}{U_j^\dag} = 0
\end{equation}
with  $K_-$ playing the role of the central charge.

The quartic invariant $\mathcal{I}_4$ of $SO(d-2)_L \times SU(1,1)_M$ subgroup becomes  a linear function of the quadratic Casimir of $SO(d-2)_L \times SU(1,1)_M$  after quantization. As a result the grade $+2$ generator $K_+$ becomes:
\begin{equation}
K_+
= \frac{1}{2} p^2
  + \frac{1}{x^2} \mathcal{G}
\end{equation}
where $\mathcal{G}$ depends on $\mathcal{L}^2$:
\begin{equation}
\mathcal{G}
= \frac{1}{4} \mathcal{L}^2 + \frac{1}{8} \left( d - 3 \right) \left( d - 5 \right)
= 2 \mathcal{M}^2 + \frac{3}{8}
\label{IsotonicCouplingConstant}
\end{equation}
The remaining $2(d-2)$ generators in grade $+1$ subspace can be obtained by taking the commutators between the  grade $-1$ generators and $K_+$:
\begin{equation}
W_i
= - i \commute{U_i}{K_+}
\qquad \qquad \qquad
W_i^\dag
= - i \commute{U_i^\dag}{K_+}
\label{SO(d,2)grade+1}
\end{equation}
Explicitly one finds
\begin{equation}
\begin{split}
W_i
&= p \, a_i - \frac{i}{x} \left[ \frac{1}{2} \left( d - 3 \right) \, a_i + i \, L_{ij} \, a_j \right]
\\
W_i^\dag
&= p \, a_i^\dag - \frac{i}{x} \left[ \frac{1}{2} \left( d - 3 \right) \, a_i^\dag + i \, L_{ij} \, a_j^\dag \right] \,.
\end{split}
\end{equation}
The grade $+2$ and grade $+1$ generators form a Heisenberg algebra as well:
\begin{equation}
\commute{W_i}{W_j^\dag} = 2 \, \delta_{ij} \, K_+
\qquad \qquad
\commute{W_i}{W_j} = \commute{W_i^\dag}{W_j^\dag} = 0
\end{equation}
with the generator $K_+$ playing the role of the central charge. The commutators of grade $-2$ and grade $+1$ generators close into grade $-1$ subspace:
\begin{equation}
\commute{W_i}{K_-} = - i \, U_i
\qquad \qquad
\commute{W_i^\dag}{K_-} = - i \, U_i^\dag
\end{equation}
Grade $\pm 2$ generators, together with the generator $\Delta$ from grade 0 subspace, form a distinguished $\mathfrak{su}(1,1)$ subalgebra, which we shall denote as $\mathfrak{su}(1,1)_K$:
\begin{equation}
\commute{K_-}{K_+} = i \, \Delta
\qquad \qquad
\commute{\Delta}{K_\pm} = \pm 2 i \, K_\pm
\end{equation}
The quadratic Casimir of  $\mathfrak{su}(1,1)_K$ subalgebra, denoted by
\begin{equation}
\mathcal{C}_2 \left[ \mathfrak{su}(1,1)_K \right]
= \mathcal{K}^2
= \Delta^2 - 2 \left( K_+ K_- + K_- K_+ \right)
\end{equation}
is related to the quadratic Casimir of $\mathfrak{so}(d-2)_L$ (and that of $\mathfrak{su}(1,1)_M$) as
\begin{equation}
\mathcal{K}^2 = - \frac{1}{2} \mathcal{L}^2 - \frac{1}{4} \left( d - 2 \right) \left( d - 6 \right) = - 4 \, \mathcal{M}^2 \,.
\end{equation}

Grade $-1$ (grade $+1$) generators transform under $SO(d-2)_L \times SU(1,1)_M$ as follows:
\begin{equation}
\begin{aligned}
\commute{M_0}{U_i} &= - \frac{1}{2} \, U_i
\\
\commute{M_+}{U_i} &= - U_i^\dag
\\
\commute{M_-}{U_i} &= 0
\\
\commute{L_{ij}}{U_k} &= i \left( \delta_{jk} U_i - \delta_{ik} U_j \right)
\end{aligned}
\qquad \qquad \qquad
\begin{aligned}
\commute{M_0}{W_i} &= - \frac{1}{2} \, W_i
\\
\commute{M_+}{W_i} &= - W_i^\dag
\\
\commute{M_-}{W_i} &= 0
\\
\commute{L_{ij}}{W_k} &= i \left( \delta_{jk} W_i - \delta_{ik} W_j \right)
\end{aligned}
\end{equation}
\begin{equation}
\begin{aligned}
\commute{M_0}{U_i^\dag} &= \frac{1}{2} \, U_i^\dag
\\
\commute{M_+}{U_i^\dag} &= 0
\\
\commute{M_-}{U_i^\dag} &= U_i
\\
\commute{L_{ij}}{U_k^\dag} &= i \left( \delta_{jk} U_i^\dag - \delta_{ik} U_j^\dag \right)
\end{aligned}
\qquad \qquad \qquad
\begin{aligned}
\commute{M_0}{W_i^\dag} &= \frac{1}{2} \, W_i^\dag
\\
\commute{M_+}{W_i^\dag} &= 0
\\
\commute{M_-}{W_i^\dag} &= W_i
\\
\commute{L_{ij}}{W_k^\dag} &= i \left( \delta_{jk} W_i^\dag - \delta_{ik} W_j^\dag \right)
\end{aligned}
\end{equation}
The  commutators between grade $-1$ generators and grade $+1$ generators are:
\begin{equation}
\begin{aligned}
\commute{U_i}{W_j}
&= 2 i \, \delta_{ij} \, M_-
\\
\commute{U_i^\dag}{W_j^\dag}
&= 2 i \, \delta_{ij} \, M_+
\end{aligned}
\qquad \qquad
\begin{aligned}
\commute{U_i^\dag}{W_j}
&= \delta_{ij} \,\left( 2 i \, M_0 - \Delta \right)
   + 2 \, L_{ij}
\\
\commute{U_i}{W_j^\dag}
&= \delta_{ij} \,\left( 2 i \, M_0 + \Delta \right)
   - 2 \, L_{ij}
\end{aligned}
\end{equation}

Let us now present  the quadratic Casimir of $\mathfrak{so}(d,2)$. One finds that the quadratic operator 
\begin{equation}
\left[ U W \right]
= U_i W_i^\dag + W_i^\dag U_i - U_i^\dag W_i - W_i U_i^\dag
\end{equation}
constructed out of the generators of grade $\pm1$ subspaces corresponding to the coset space
\begin{equation*}
\frac{SO(d,2)}{SO(d-2) \times SU(1,1)} \,,
\end{equation*}
is related to the quadratic Casimir of $\mathfrak{su}(1,1)_K$ as follows:
\begin{equation}
\left[ UW \right] = - 4 i \, \mathcal{K}^2 + 4 i \left( d -  2 \right)
\end{equation}
As a consequence one finds\footnote{That there exists a two-parameter family of quadratic invariants for the minrep was shown by Joseph \cite{Joseph:1974,joseph1976minimal}.}:
\begin{equation}
\begin{split}
&\mathcal{C}_2 \left[ \mathfrak{so}(d-2)_L \right]
+ \lambda_1 \, \mathcal{C}_2 \left[ \mathfrak{su}(1,1)_M \right]
+ \lambda_2 \, \mathcal{C}_2 \left[ \mathfrak{su}(1,1)_K \right]
+ \frac{i}{2} \left( 1 + \frac{\lambda_1}{8} - \frac{\lambda_2}{2} \right) \left[ UW \right]
\\
& \qquad
= - \frac{1}{4} \left( d - 2 \right) \lambda_1 + \left( d - 2 \right) \lambda_2 - \frac{1}{2} \left( d - 2 \right)^2
\end{split}
\end{equation}
The quadratic Casimir of $\mathfrak{so}(d,2)$ is given by  $\lambda_1 = 4$ and $\lambda_2 = - 1$ (for any $d$) and is therefore given by
\begin{equation}
\mathcal{C}_2 \left[ \mathfrak{so}(d,2) \right] = - \frac{1}{2} \left( d^2 -  4 \right) \,.
\end{equation}

%%%%%%%%%%%%%%%%%%%%%%%%%%%%%%%%%%%%%%%%%%%%%%%%%
%%%%% SO(d,2) Noncompact 3-Grading %%%%%%%%%%%%%%%%%%%%%%%%%%%%
%%%%%%%%%%%%%%%%%%%%%%%%%%%%%%%%%%%%%%%%%%%%%%%%%

\section{Noncompact 3-grading of $\mathfrak{so}(d,2)$ with respect to the subalgebra $\mathfrak{so}(d-1,1) \oplus \mathfrak{so}(1,1)$}
\label{sec:SO(d,2)nc3G}

Considered as the $d$-dimensional conformal group, $SO(d,2)$ has a natural 3-grading with respect to the generator $\mathcal{D}$ of dilatations, whose eigenvalues determine the conformal dimensions of operators and states. We shall denote the corresponding 3-graded decomposition of $\mathfrak{so}(d,2)$ as
\begin{equation}
\mathfrak{so}(d,2)
= \mathfrak{N}^- \oplus
  \mathfrak{N}^0 \oplus
  \mathfrak{N}^+
\end{equation}
which symbolically satisfy
\begin{equation}
\commute{\mathcal{D}}{\mathfrak{N}^0} = 0
\qquad \qquad
\commute{\mathcal{D}}{\mathfrak{N}^+} = + i \, \mathfrak{N}^+
\qquad \qquad
\commute{\mathcal{D}}{\mathfrak{N}^-} = - i \, \mathfrak{N}^-
\end{equation}
with $\mathfrak{N}^0 = \mathfrak{so}(d-1,1) \oplus \mathfrak{so}(1,1)_{\mathcal{D}}$. The subalgebra $\mathfrak{so}(d-1,1)$ in $\mathfrak{N}^{0}$ represents the Lorentz algebra in $d$ dimensions and grade $+1$ and $-1$ subspaces are spanned by translation and special conformal generators, respectively.

The \emph{noncompact} dilatation generator $\mathfrak{so}(1,1)_{\mathcal{D}}$ is given by
\begin{equation}
\mathcal{D}
= \frac{1}{2} \left[ \Delta - i \left( M_+ - M_- \right) \right] \,.
\label{Dilatation}
\end{equation}
The Lorentz group generators $\mathcal{M}_{\mu\nu}$ ($\mu,\nu = 0,\dots,d-1$) are given by
\begin{equation}
\begin{split}
\mathcal{M}_{0i}
= \frac{1}{2\sqrt{2}} \left( U_i + U_i^\dag \right)
  + \frac{i}{2\sqrt{2}} \left( W_i - W_i^\dag \right)
& \qquad
\mathcal{M}_{ij}
= L_{ij}
\\
\mathcal{M}_{i,d-1}
= \frac{1}{2\sqrt{2}} \left( U_i + U_i^\dag \right)
  - \frac{i}{2\sqrt{2}} \left( W_i - W_i^\dag \right)
& \qquad
\mathcal{M}_{0,d-1}
= \frac{1}{2} \left[ \Delta + i \left( M_+ - M_- \right) \right]
\end{split}
\label{lorentz}
\end{equation}
and satisfy the commutation relations:
\begin{equation}
\commute{\mathcal{M}_{\mu\nu}}{\mathcal{M}_{\rho\tau}}
= i \left(
     \eta_{\nu\rho} \mathcal{M}_{\mu\tau}
     - \eta_{\mu\rho} \mathcal{M}_{\nu\tau}
     - \eta_{\nu\tau} \mathcal{M}_{\mu\rho}
     + \eta_{\mu\tau} \mathcal{M}_{\nu\rho}
    \right)
\end{equation}
where $\eta_{\mu\nu} = \mathrm{diag} (-,+,\dots,+)$.

The translation generators $\mathcal{P}_\mu$ ($\mu = 0,\dots,d-1$) are given by
\begin{equation}
\begin{split}
\mathcal{P}_0
&= K_+ + M_0 + \frac{1}{2} \left( M_+ + M_- \right)
\\
\mathcal{P}_i
&= \frac{1}{\sqrt{2}} \left( W_i + W_i^\dag \right) \qquad \qquad (i = 1,\dots,d-2)
\\
\mathcal{P}_{d-1}
&= K_+ - M_0 - \frac{1}{2} \left( M_+ + M_- \right)
\end{split}
\end{equation}
and the special conformal generators $\mathcal{K}_\mu$ ($\mu = 0,\dots,d-1$) are given by
\begin{equation}
\begin{split}
\mathcal{K}_0
&= K_- + M_0 - \frac{1}{2} \left( M_+ + M_- \right)
\\
\mathcal{K}_i
&= - \frac{i}{\sqrt{2}} \left( U_i - U_i^\dag \right) \qquad \qquad (i = 1,\dots,d-2)
\\
\mathcal{K}_{d-1}
&= - K_- + M_0 - \frac{1}{2} \left( M_+ + M_- \right) \,.
\end{split}
\end{equation}

They satisfy the commutation relations:
\begin{equation}
\begin{split}
\commute{\mathcal{M}_{\mu\nu}}{\mathcal{M}_{\rho\tau}}
&= i \left(
     \eta_{\nu\rho} \mathcal{M}_{\mu\tau}
     - \eta_{\mu\rho} \mathcal{M}_{\nu\tau}
     - \eta_{\nu\tau} \mathcal{M}_{\mu\rho}
     + \eta_{\mu\tau} \mathcal{M}_{\nu\rho}
    \right)
\\
\commute{\mathcal{P}_\mu}{\mathcal{M}_{\nu\rho}}
&= i \left( \eta_{\mu\nu} \, \mathcal{P}_\rho
            - \eta_{\mu\rho} \, \mathcal{P}_\nu
     \right)
\\
\commute{\mathcal{K}_\mu}{\mathcal{M}_{\nu\rho}}
&= i \left( \eta_{\mu\nu} \, \mathcal{K}_\rho
            - \eta_{\mu\rho} \, \mathcal{K}_\nu
     \right)
\\
\commute{\mathcal{D}}{\mathcal{M}_{\mu\nu}}
&= \commute{\mathcal{P}_\mu}{\mathcal{P}_\nu}
 = \commute{\mathcal{K}_\mu}{\mathcal{K}_\nu}
 = 0
\\
\commute{\mathcal{D}}{\mathcal{P}_\mu}
&= + i \, \mathcal{P}_\mu
\qquad \qquad
\commute{\mathcal{D}}{\mathcal{K}_\mu}
 = - i \, \mathcal{K}_\mu
\\
\commute{\mathcal{P}_\mu}{\mathcal{K}_\nu}
&= 2 i \left( \eta_{\mu\nu} \, \mathcal{D} + \mathcal{M}_{\mu\nu} \right)
\end{split}
\end{equation}

Finally we note that the Poincar\'e mass operator in $d$ dimensions vanishes identically
\begin{equation}
\mathscr{M}^2 = \eta_{\mu\nu} \mathcal{P}^\mu\mathcal{P}^\nu
              = 0
\end{equation}
for the minimal unitary realization given above. Therefore the minimal unitary representation of $SO(d,2)$ corresponds to a massless representation in $d$ dimensions. In addition we  have
\begin{equation}
\eta^{\mu\nu} \mathcal{K}_\mu \mathcal{K}_\nu = 0 \,.
\end{equation}

%%%%%%%%%%%%%%%%%%%%%%%%%%%%%%%%%%%%%%%%%%%%%%%%%
%%%%% SO(d,2) Compact 3-Grading %%%%%%%%%%%%%%%%%%%%%%%%%%%%%%
%%%%%%%%%%%%%%%%%%%%%%%%%%%%%%%%%%%%%%%%%%%%%%%%%

\section{Compact 3-grading of $\mathfrak{so}(d,2)$ with respect to the subalgebra $\mathfrak{so}(d) \oplus \mathfrak{so}(2) $}
\label{sec:SO(d,2)c3G}

The Lie algebra $\mathfrak{so}(d,2)$ has a 3-grading:
\begin{equation}
\mathfrak{so}(d,2)
= \mathfrak{C}^- \oplus
  \mathfrak{C}^0 \oplus
   \mathfrak{C}^+ \,.
\end{equation}
with respect to its maximal compact subalgebra $\mathfrak{C}^0 = \mathfrak{so}(d) \oplus \mathfrak{so}(2)$, determined by the $\mathfrak{u}(1)$ generator
\begin{equation}
H = \frac{1}{2} \left( K_+ + K_- \right) + M_0 \,.
\label{SO(2)generator}
\end{equation}

The operators in various grade subspaces satisfy
\begin{equation}
\commute{H}{\mathfrak{C}^0} = 0
\qquad \qquad
\commute{H}{\mathfrak{C}^+} = + \, \mathfrak{C}^+
\qquad \qquad
\commute{H}{\mathfrak{C}^-} = - \, \mathfrak{C}^- \,.
\end{equation}

The $\mathfrak{so}(d)$ generators $\widetilde{M}_{MN}$ ($M,N = 1,\dots,d$) in grade 0 subspace $\mathfrak{C}^0$ are given by
\begin{equation}
\begin{aligned}
\widetilde{M}_{ij}
&= L_{ij}
\\
\widetilde{M}_{d-1,d}
&=\frac{1}{2} \left( \mathcal{P}_{d-1} - \mathcal{K}_{d-1} \right)
\end{aligned}
\qquad \qquad
\begin{aligned}
\widetilde{M}_{i,d-1}
&= \mathcal{M}_{i,d-1} \\
\widetilde{M}_{i,d}
&=\frac{1}{2} \left( \mathcal{P}_i - \mathcal{K}_i \right)
\end{aligned}
\label{SO(d)generators}
\end{equation}
and satisfy the commutation relations
\begin{equation}
\commute{\widetilde{M}_{MN}}{\widetilde{M}_{PQ}}
= i \left(
     \delta_{NP} \widetilde{M}_{MQ} - \delta_{MP} \widetilde{M}_{NQ}
     - \delta_{NQ} \widetilde{M}_{MP} + \delta_{MQ} \widetilde{M}_{NP}
    \right) \,.
\end{equation}
The generators $\widetilde{M}_{ij} \oplus \widetilde{M}_{d-1,d}$ form the $\mathfrak{so}(d-2)_L \oplus \mathfrak{so}(2)$ subalgebra of $\mathfrak{so}(d) $.

We shall label the $d$ operators that belong to grade $+1$ subspace $\mathfrak{C}^+$ as $\widetilde{B}_M^\dag$ ($M = 1,\dots,d$) where
\begin{equation}
\begin{split}
\widetilde{B}_i^\dag
&= \frac{1}{\sqrt{2}} \left( U_i^\dag - i \, W_i^\dag \right)
\qquad \qquad (i = 1,\dots,d-2)
\\
\widetilde{B}_{d-1}^\dag
&= \frac{1}{2} \left[ \Delta - i \left( K_+ - K_- \right) \right] + i \, M_+
\\
\widetilde{B}_d^\dag
&= \frac{i}{2} \left[ \Delta - i \left( K_+ - K_- \right) \right] + M_+ \,.
\end{split}
\label{SO(d,2)3gradingC+}
\end{equation}
These operators in grade $+1$ subspace $\mathfrak{C}^+$ satisfy the following important relation:
\begin{equation}
\widetilde{B}_M^\dag \widetilde{B}_M^\dag
= \widetilde{B}_1^\dag \widetilde{B}_1^\dag
  + \widetilde{B}_2^\dag \widetilde{B}_2^\dag
  + \dots
  + \widetilde{B}_d^\dag \widetilde{B}_d^\dag
= 0
\label{MinrepC+Constraint}
\end{equation}
which corresponds to the masslessness condition in the noncompact picture.

Similarly, we shall label the $d$ operators that belong to grade $-1$ subspace $\mathfrak{C}^-$ as $\widetilde{B}_M$ ($M,N,\dots = 1,\dots,d$) where
\begin{equation}
\begin{split}
\widetilde{B}_i
&= \frac{1}{\sqrt{2}} \left( U_i + i \, W_i \right)
\qquad \qquad (i = 1,\dots,d-2)
\\
\widetilde{B}_{d-1}
&= \frac{1}{2} \left[ \Delta + i \left( K_+ - K_- \right) \right] - i \, M_-
\\
\widetilde{B}_d
&= - \frac{i}{2} \left[ \Delta + i \left( K_+ - K_- \right) \right] + M_- \,.
\end{split}
\label{SO(d,2)3gradingC-}
\end{equation}

The commutation relations of the $SO(d,2)$ generators in this compact basis can be listed as:
\begin{equation}
\begin{split}
\commute{\widetilde{M}_{MN}}{\widetilde{M}_{PQ}}
&= i \left(
      \delta_{NP} \widetilde{M}_{MQ} - \delta_{MP} \widetilde{M}_{NQ}
      - \delta_{NQ} \widetilde{M}_{MP} + \delta_{MQ} \widetilde{M}_{NP}
     \right)
\\
\commute{\widetilde{B}_M^\dag}{\widetilde{M}_{NP}}
&= i \left(
      \delta_{MN} \, \widetilde{B}_P^\dag - \delta_{MP} \, \widetilde{B}_N^\dag
     \right)
\\
\commute{\widetilde{B}_M}{\widetilde{M}_{NP}}
&= i \left(
      \delta_{MN} \, \widetilde{B}_P - \delta_{MP} \, \widetilde{B}_N
     \right)
\\
\commute{H}{\widetilde{M}_{MN}}
&= \commute{\widetilde{B}_M^\dag}{\widetilde{B}_N^\dag}
 = \commute{\widetilde{B}_M}{\widetilde{B}_N}
 = 0
\\
\commute{H}{\widetilde{B}_M^\dag}
&= + \widetilde{B}_M^\dag
\qquad \qquad
\commute{H}{\widetilde{B}_M}
 = - \widetilde{B}_M
\\
\commute{\widetilde{B}_M^\dag}{\widetilde{B}_N}
&= 2  \left( - \delta_{MN} \, H + i \, \widetilde{M}_{MN} \right)
\end{split}
\label{SO(d,2)Compact}
\end{equation}

The operators in the subspace $\mathfrak{C}^+$ given in equation (\ref{SO(d,2)3gradingC+}) are the Hermitian conjugates of those in the subspace $\mathfrak{C}^-$ given in equation (\ref{SO(d,2)3gradingC-}). The $\mathfrak{so}(2)$ generator $H$ in the subspace $\mathfrak{C}^0$ is simply the conformal Hamiltonian or the $AdS$ energy. In terms of the generators in noncompact 3-grading of $\mathfrak{so}(d,2)$  this conformal Hamiltonian (or the $AdS$ energy operator) is simply  $\frac{1}{2} \left( \mathcal{P}_0 + \mathcal{K}_0 \right)$.

%%%%%%%%%%%%%%%%%%%%%%%%%%%%%%%%%%%%%%%%%%%%%%%%%
%%%%%% SU(1,1)_K of SO(d,2) %%%%%%%%%%%%%%%%%%%%%%%%%%%%%%%%
%%%%%%%%%%%%%%%%%%%%%%%%%%%%%%%%%%%%%%%%%%%%%%%%%

\section{Distinguished $SU(1,1)_K$ subgroup of $SO(d,2)$ generated by the isotonic (singular) oscillators}
\label{sec:SU(1,1)_K}

The $\mathfrak{u}(1)$ generator $H$, given in equation (\ref{SO(2)generator}), can be expressed in terms of the $a$-type bosonic oscillators, singlet coordinate $x$, and its conjugate momentum $p$ in the following form:
\begin{equation}
\begin{split}
H &= \frac{1}{2} \left( K_+ + K_- \right) + M_0
\\
 &= \frac{1}{4} \left( x^2 + p^2 \right)
     + \frac{1}{2 \, x^2} \mathcal{G}
     + \frac{1}{2} \, a_i^\dag a_i
     + \frac{1}{4} \left( d - 2 \right)
\\
  &= H_\odot + H_a
\end{split}
\end{equation}
where
\begin{equation}
H_\odot
= \frac{1}{4} \left( x^2 + p^2 \right) + \frac{1}{2 \, x^2} \mathcal{G}
\qquad \qquad
H_a=  M_0
= \frac{1}{2} \, a_i^\dag a_i + \frac{1}{4} \left( d - 2 \right) \,.
\end{equation}
As pointed out earlier, this $\mathfrak{u}(1)$ generator $H$ is the $d$-dimensional conformal Hamiltonian or the $(d+1)$-dimensional $AdS$ energy operator, depending on whether one is treating $SO(d,2)$ as the $d$-dimensional conformal group or as the $(d+1)$-dimensional $AdS$ group. The part $H_a$  is simply 1/2 the Hamiltonian of standard bosonic oscillators $a_i$ and the part $H_\odot$ is 1/2 the Hamiltonian of a singular harmonic oscillator with a potential function
\begin{equation}
V \left( x \right) = \frac{\mathcal{G}}{x^2}
\end{equation}
where $\mathcal{G}$ was given in equation (\ref{IsotonicCouplingConstant}).  We note that  $H_\odot$  corresponds to the Hamiltonian in the conformal quantum mechanics of  \cite{deAlfaro:1976je}, with the operator $\mathcal{G}$ replacing  the ``coupling constant'' \cite{Gunaydin:2001bt}. It also appears in this form in the Hamiltonian of the Calogero models  \cite{Calogero:1969af,Calogero:1970nt} as well as in isotonic singular oscillator models  \cite{Casahorran:1995vt,MR2451306}.

Since the singlet coordinate $x$ and its conjugate momentum $p$ enter the minrep as in conformal quantum mechanics, we shall use the coordinate (Schr\"odinger) picture for the states that  form the basis of an irrep of the subgroup $SU(1,1)_K$. Consider now this conformal quantum mechanics  Hamiltonian:
\begin{equation}
\begin{split}
H_\odot
 = \frac{1}{2} \left( K_+ + K_- \right)
&= \frac{1}{4} \left( x^2 + p^2 \right)
   + \frac{1}{2 \, x^2} \mathcal{G}
\\
&= \frac{1}{4} \left( x^2 - \frac{\partial^2}{\partial x^2} \right)
   + \frac{1}{2 \, x^2} \mathcal{G} 
\end{split}
\label{SingularHamiltonian}
\end{equation}
Together with the generators $\widetilde{B}_{d-1}^\dag$ and  $\widetilde{B}_d^\dag$ it generates the distinguished $SU(1,1)_K$ subalgebra. In the compact 3-grading determined by $H_\odot$,  the noncompact  generators take the form
\begin{equation}
\begin{split}
\widetilde{B}_\odot^\dag
 = - \frac{i}{2} \left( \widetilde{B}_{d-1}^\dag - i \, \widetilde{B}_d^\dag \right)
 = - \frac{i}{2} \left[ \Delta - i \left( K_+ - K_- \right) \right]
&= \frac{1}{4} \left( x - i p \right)^2
   - \frac{1}{2 \, x^2} \mathcal{G}
\\
 &= \frac{1}{4} \left( x - \frac{\partial}{\partial x} \right)^2
    - \frac{1}{2 \, x^2} \mathcal{G}
\\
\widetilde{B}_\odot
 = \frac{i}{2} \left( \widetilde{B}_{d-1} + i \, \widetilde{B}_d \right)
 = \frac{i}{2} \left[ \Delta + i \left( K_+ - K_- \right) \right]
&= \frac{1}{4} \left( x + i p \right)^2
   - \frac{1}{2 \, x^2} \mathcal{G}
\\
&= \frac{1}{4} \left( x + \frac{\partial}{\partial x} \right)^2
   - \frac{1}{2 \, x^2} \mathcal{G}
\end{split}
\end{equation}
and satisfy the commutation relations:
\begin{equation}
\commute{\widetilde{B}_\odot}{\widetilde{B}_\odot^\dag} = 2 \, H_\odot
\qquad \qquad
\commute{H_\odot}{\widetilde{B}_\odot^\dag} = + \, \widetilde{B}_\odot^\dag
\qquad \qquad
\commute{H_\odot}{\widetilde{B}_\odot} = - \, \widetilde{B}_\odot
\end{equation}

Now we consider the Fock space of the $a$-type oscillators spanned by 
the states of the form
\begin{equation}
\ket{n_1,n_2,\dots,n_{d-2}}
= \prod_i \frac{1}{\sqrt{n_{i} !}} \, ( \, a_i^\dag \, )^{n_{i}} \ket{0}
\end{equation}
where $n_{i}$ are non-negative integers and 
 the vacuum state $\ket{0}$  is annihilated by all bosonic annihilation operators $a_i$:
\begin{equation}
a_i \ket{0} = 0
\qquad \qquad
\left( i = 1,\dots,d-2 \right)
\end{equation}

The state(s) with the lowest $H_\odot$ eigenvalue are obtained by taking the tensor product of states $\ket{\Lambda_g}$ with the lowest eigenvalue $g$ of $\mathcal{G}$ and the lowest weight vector of $SU(1,1)_K$ determined by $g$ \cite{MR858831}:
 \be \psi_0^{\alpha_g} \left( x \right)
= C_0 \, x^{\alpha_g} e^{-x^2/2} \ee
where $C_0$ is a normalization constant and
\begin{equation}
\alpha_g
= \frac{1}{2} \pm \sqrt{2 g + \frac{1}{4}} \,.
\end{equation}
These states satisfy
\begin{equation}
\widetilde{B}_\odot \, \psi_0^{\alpha_g} \left( x \right) \ket{\Lambda_g}
= 0 \,.
\end{equation}
The Hermiticity of $H_\odot$ requires that
\begin{equation}
g \geq - \frac{1}{8}
\end{equation}
and the normalizability of the lowest weight vector  requires
\begin{equation}
\alpha_g > - \frac{1}{2} \,.
\end{equation}
The state $\psi_0^{\alpha_g} \left( x \right) \ket{\Lambda_g}$ is an eigenstate of $H_\odot$ with eigenvalue
\begin{equation}
E_{\odot,0}^{\alpha_g}
= \frac{\alpha_g}{2} + \frac{1}{4}
= \frac{1}{2} \pm \frac{1}{2} \sqrt{2 g + \frac{1}{4}} \,.
\label{alpha_g}
\end{equation}

For the minrep of $SO(d,2)$ given earlier, the lowest possible value of $g$ is
\begin{equation}
g_0 = \frac{1}{8} \left( d - 3 \right) \left( d - 5 \right)
\end{equation}
which occurs when $\ket{\Lambda_g}$ is simply the Fock vacuum $\ket{0}$ of $a$-type oscillators. For this $g$ we have two possible values of $\alpha_g$, namely $\left( 5 - d \right) / 2$ and $\left( d - 3 \right) / 2$. For all $d \geq 6$, the state with $\alpha_g = \left( 5 - d \right) / 2$ is non-normalizable. For $d=4$, $\alpha_g = \left( 5 - d \right) / 2$ produces the same result as $\alpha_g = \left( d - 3 \right) / 2$. For $d=5$, even though the state with $\alpha_g = \left( 5 - d \right) / 2$ is normalizable, it leads to non-normalizable states under the action of $ SO(5)$ when $SU(1,1)_K$ is extended to $SO(5,2)$ \cite{Fernando:2014pya}. Therefore we need to choose
\begin{equation}
\alpha_g = \frac{(d-3)}{2} \,.
\end{equation}
The corresponding tensor product state
\begin{equation}
\psi_0^{\alpha_g} \left( x \right) \, \ket{0}
= C_0 \, x^{(d-3)/2} \, e^{-x^2/2} \ket{0}
\end{equation}
is an eigenstate of $H_\odot$ with the lowest eigenvalue
\begin{equation}
E_{\odot,0}^{\alpha_g} = \frac{1}{4} \left( d - 2 \right) \,.
\end{equation}
 The higher  eigenstates of $H_\odot$ can be obtained from this ``ground state'' by repeatedly acting on it with the raising operator $\widetilde{B}_\odot^\dag$:
\begin{equation}
| \, \psi^{\alpha_g}_n \rangle \equiv \psi_n^{\alpha_g} \left( x \right) \, \ket{0}
 = C_n \, ( \widetilde{B}_\odot^\dag )^n \,
   \psi_0^{\alpha_g} \left( x \right) \ket{0} \qquad \qquad (n=0,1,2,...)
\label{isotonicgroundstate}
\end{equation}
where $C_n$ are normalization constants. They satisfy
\begin{equation}
H_\odot \psi_n^{\alpha_g} \left( x \right) \, \ket{0}=
E_{\odot,n}^{\alpha_g} \psi_n^{\alpha_g} \left( x \right) \, \ket{0}
\end{equation}
where
\begin{equation}
E_{\odot,n}^{\alpha_g}= E_{\odot,0}^{\alpha_g} + n
 = \frac{1}{4} \left( d - 2 \right) + n \,.
\end{equation}
The states $| \, \psi^{\alpha_g}_n \, \rangle$ form the  basis of a unitary lowest weight representation  of $SU(1,1)_K$.

%%%%%%%%%%%%%%%%%%%%%%%%%%%%%%%%%%%%%%%%%%%%%%%%%
%%%%%% K-Type decomposition %%%%%%%%%%%%%%%%%%%%%%%%%%%%%%%
%%%%%%%%%%%%%%%%%%%%%%%%%%%%%%%%%%%%%%%%%%%%%%%%%

\section{$K$-type decomposition of the minimal unitary representation  of $SO(d,2)$ }
\label{sec:undeformedminrep}

The K-type decomposition of the minimal unitary representation of $SO(d,2)$ is simply the decomposition with respect to its maximal compact subgroup $SO(d) \times SO(2)_H$.  For the true minrep of $SO(d,2)$ this turns out to be easily determined since the lowest energy representation is an $SO(d)$ singlet, namely the state
\begin{equation}
 | \psi_0^{\alpha_g} \rangle
= C_0 \, x^{(d-3)/2} \, e^{-x^2/2} \ket{0}
\end{equation}
 with the lowest eigenvalue  $E = \frac{1}{2} \left( d - 2 \right)$ of  the $\mathfrak{so}(2)$ generator $H$. This state $ | \psi_0^{\alpha_g} \rangle$ is annihilated by all the operators $\widetilde{B}_1,\dots,\widetilde{B}_d$ in the subspace $\mathfrak{C}^-$ of $\mathfrak{so}(d,2)$ given in equation (\ref{SO(d,2)3gradingC-}):
\begin{equation}
\widetilde{B}_M  | \psi_0^{\alpha_g} \rangle = 0
\qquad \qquad (M=1,\dots,d)
\end{equation}
Therefore the minrep of $SO(d,2)$ is a unitary lowest weight representation.
The irreducible unitary lowest weight representations of a non-compact group are uniquely labelled by their lowest weight vectors. For $SO(d,2)$ these lowest weight vectors have the lowest eigenvalue of the generator $H$ of $SO(2)$ subgroup corresponding to the $AdS_{(d+1)}$ energy or of the $d$-dimensional conformal Hamiltonian. By acting on this lowest weight vector with the generators of $SO(d)$ one generates the lowest energy irrep of $SO(d)\times SO(2)$ which we will refer to as the lowest (energy) K-type. For physical applications it is more convenient to use the labels of the lowest K-type to uniquely designate an irreducible unitary lowest weight representation. Therefore we will  label the lowest (energy) K-type as
\begin{equation}
| \, E_0 \,;\, (n_1,n_2,\dots,n_p) \, \rangle_0
\end{equation}
where $E_0$ indicates the eigenvalue of the $SO(2)$ generator $H$
\begin{equation}
H \, | \, E_0 \,;\, (n_1,n_2,\dots,n_p) \, \rangle_0 = E_0 \, | \, E_0 \, ; \, (n_1,n_2,\dots,n_p) \, \rangle_0 \,,
\end{equation}
$(n_1,n_2,\dots,n_p)$ are the Dynkin labels of the irrep of lowest K-type under $SO(d)$, and $p=[\frac{d}{2}]$.

Therefore the lowest K-type of the minrep of $SO(d,2)$ is the $SO(d)$ singlet state $| \psi_0^{\alpha_g} \rangle $, which will be denotes as
\begin{equation}
| \psi_0^{\alpha_g} \rangle  = | \, \frac{1}{2}(d-2) \,;\, (0,0,\dots,0) \, \rangle_0 \,.
\end{equation}
All the other states of the particle basis of the minrep with higher energies can be obtained from the lowest K-type by acting on it repeatedly with the operators $\widetilde{B}_1^\dag,\dots,\widetilde{B}_d^\dag$ in the subspace $\mathfrak{C}^+$ of $\mathfrak{so}(d,2)$ given in equation (\ref{SO(d,2)3gradingC+}):\footnote{The round brackets with a subscript $o$  in $B_{(M_1}^\dag \dots B_{M_n)_o}^\dag$ denote the symmetric traceless product of the operators.}
\begin{equation}
| \psi_0^{\alpha_g} \rangle
\quad , \quad
\widetilde{B}_{M_1}^\dag | \psi_0^{\alpha_g} \rangle
\quad , \quad
\widetilde{B}_{(M_1}^\dag \widetilde{B}_{M_2)_o}^\dag | \psi_0^{\alpha_g} \rangle
\quad , \quad
\widetilde{B}_{(M_1}^\dag \widetilde{B}_{M_2}^\dag \widetilde{B}_{M_3)_o}^\dag| \psi_0^{\alpha_g} \rangle
\quad , \quad
\dots\dots
\end{equation}

All the states that belong to a given energy level form a single irrep of $SO(d)$. In Table \ref{Table:Minrep}, we give the (K-type) decomposition of the minrep of $SO(d,2)$ with respect to its maximal compact subgroup and the dimension of their $SO(d)$ irrep as well as the energy. For \emph{odd} $d$ the minrep of $SO(d,2)$ is the scalar singleton representation of $SO(d,2)$ in $AdS_{(d+1)}$, just like the Dirac scalar singleton of $SO(3,2)$ in $AdS_4$. For \emph{even} $d$ it corresponds to the scalar doubleton of $SO(d,2)$ in $AdS_{(d+1)}$ as defined and studied for $d=4$ in \cite{Gunaydin:1984fk,Fernando:2009fq} and for $d=6$ in \cite{Gunaydin:1984wc,Fernando:2010dp,Fernando:2010ia}.
%\newpage

%%%%%%%%%%%%%%%%%%%%%%%%%%%%%%%%%%%%%%%%%%%%%%%%
\begin{small}
\begin{longtable}[c]{|c|c|c|c|}
\kill
%%%%%%%%%%%%%%%%%%%%%%%%%%%%%%%%%%%%%%%%%%%%%%%%

\caption[The minimal unitary representation of $SO(d,2)$]
{K-type decomposition of the minrep of $SO(d,2)$ with the lowest weight vector  $| \psi_0^{\alpha_g} \rangle = | \, \frac{1}{2}(d-2) \,;\, (0,0,\dots,0) \, \rangle_0$. The $AdS_{(d+1)}$ energy (or the eigenvalue of the $d$-dimensional conformal Hamiltonian), and the dimension and  Dynkin labels of $SO(d)$ irrep at each level are given.
\label{Table:Minrep}} \\
\hline
 & & & \\
States & $AdS_{(d+1)}$ & Dim of & Dynkin \\
 & Energy & $SO(d)$ irrep & labels of \\
 & $E$ & & $SO(d)$ irrep \\
 & & & \\
\hline
 & & & \\
\endfirsthead
%%%%%%%%%%%%%%%%%%%%%%%%%%%%%%%%%%%%%%%%%%%%%%%%
\caption[]{(continued)} \\
\hline
 & & & \\
States & $AdS_{(d+1)}$ & Dim of & Dynkin \\
 & Energy & $SO(d)$ irrep & labels of \\
 & $E$ & & $SO(d)$ irrep \\
 & & & \\
\hline
\endhead
%%%%%%%%%%%%%%%%%%%%%%%%%%%%%%%%%%%%%%%%%%%%%%%%
 & & & \\
\hline
\endlastfoot
%%%%%%%%%%%%%%%%%%%%%%%%%%%%%%%%%%%%%%%%%%%%%%%%

$| \psi_0^{\alpha_g} \rangle$ &
$\frac{d}{2} - 1$ &
1 &
$(0,0,\dots,0)$
\\[8pt]

\hline
 & & & \\

$\widetilde{B}_{M_1}^\dag | \psi_0^{\alpha_g} \rangle$ &
$\frac{d}{2}$ &
$d$ &
$(1,0,\dots,0)$
\\[8pt]

\hline
 & & & \\

$\widetilde{B}_{(M_1}^\dag \widetilde{B}_{M_2)_o}^\dag | \psi_0^{\alpha_g} \rangle$ &
$\frac{d}{2}+1$ &
$\frac{(d-1)(d+2)}{2}$ &
$(2,0,\dots,0)$
\\[8pt]

\hline
 & & & \\

$\widetilde{B}_{(M_1}^\dag \widetilde{B}_{M_2}^\dag
   \widetilde{B}_{M_3)_o}^\dag | \psi_0^{\alpha_g} \rangle$ &
$\frac{d}{2}+2$ &
$\frac{(d-1)d(d+4)}{6}$ &
$(3,0,\dots,0)$
\\[8pt]

\hline
 & & & \\

\vdots & \vdots & \vdots & \vdots \\ [8pt]

\hline
 & & & \\

$\widetilde{B}_{(M_1}^\dag \dots \widetilde{B}_{M_n)_o}^\dag
   | \psi_0^{\alpha_g} \rangle$ &
$\frac{d}{2}+n-1$ &
$\frac{(n+1)(n+2)\dots(n+d-3)(2n+d-2)}{d!}$ &
$(n,0,\dots,0)$
\\[8pt]

\hline
 & & & \\

\vdots & \vdots & \vdots & \vdots \\[8pt]

%%%%%%%%%%%%%%%%%%%%%%%%%%%%%%%%%%%%%%%%%%%%%%%
\end{longtable}
\end{small}
%%%%%%%%%%%%%%%%%%%%%%%%%%%%%%%%%%%%%%%%%%%%%%%

The  products $\widetilde{B}_{(M_1}^\dag \dots \widetilde{B}_{M_n)}^\dag$ of the operators $B_M^\dag$ from the subspace $\mathfrak{C}^+$ are symmetric and traceless due to the fact that they satisfy the condition $\widetilde{B}_M^\dag \widetilde{B}_M^\dag = 0$. As a consequence, by acting on the lowest weight vector $| \psi_0^{\alpha_g} \rangle$, which is a singlet of $SO(d)$, they generate  states whose  $SO(d)$ Young tableaux have one row of $n$ boxes:
\begin{equation}
\underbrace{\begin{picture}(100,20)(0,0)
\put(0,5){\line(0,1){10}}
\put(0,5){\line(1,0){100}}
\put(0,15){\line(1,0){100}}
\put(10,5){\line(0,1){10}}
\put(20,5){\line(0,1){10}}
\put(30,5){\line(0,1){10}}
\put(40,5){\line(0,1){10}}
\put(70,5){\line(0,1){10}}
\put(80,5){\line(0,1){10}}
\put(90,5){\line(0,1){10}}
\put(100,5){\line(0,1){10}}
\put(55,10){\makebox(0,0){$\cdots$}}
\end{picture}}_{\mbox{$n$ boxes}}
\end{equation}

%%%%%%%%%%%%%%%%%%%%%%%%%%%%%%%%%%%%%%%%%%%%%%%%%
%%%%% SO(d,2) With Deformations %%%%%%%%%%%%%%%%%%%%%%%%%%%%%%
%%%%%%%%%%%%%%%%%%%%%%%%%%%%%%%%%%%%%%%%%%%%%%%%%

\section{Deformations of the minimal unitary representation of $SO(d,2)$}
\label{sec:deformedSO(d,2)}

In the previous section, we studied the minrep of $SO(d,2)$ using quasiconformal techniques and showed that it describes a massless scalar conformal field in $d$ dimensions. In this section we shall describe how to obtain all possible ``deformations'' of the minrep of $SO(d,2)$ by extending the little group of massless states $SO(d-2)_L$ generated by the ``orbital'' operators $L_{ij}$ by the addition of ``spin operators'' $S_{ij}$:
\begin{equation}
L_{ij} \longrightarrow J_{ij}= L_{ij} + S_{ij}
\label{SO(d)J}
\end{equation}
We shall denote the subgroup generated by $J_{ij}$ as $SO(d-2)_J$ and its quadratic Casimir as $\mathcal{J}^2$:
\begin{equation}
\mathcal{C}_2 \left[ \mathfrak{so}(d-2)_J \right]
= \mathcal{J}^2
= J_{ij} J_{ij}
= \mathcal{L}^2 + \mathcal{S}^2 + 2\, \mathcal{L} \cdot \mathcal{S}
\end{equation}
where $\mathcal{S}^2$ denotes the quadratic Casimir of $\mathfrak{so}(d-2)_S$ generated by $S_{ij}$ and $\mathcal{L} \cdot \mathcal{S} = L_{ij} S_{ij}$.

With the inclusion of spin terms, one can write a general Ansatz for $K_+$, which is invariant under $SO(d-2)_J$, and impose the constraints due to Jacobi identities. Then one finds that all the Jacobi identities are satisfied if $K_+$ has the form
\begin{equation}
K_+ = \frac{1}{2} p^2
      + \frac{1}{x^2} \left(
                            \frac{1}{2} \, \mathcal{J}^2 - \frac{1}{4} \, \mathcal{L}^2
                            - \frac{(d-6)}{2(d-2)} \, \mathcal{S}^2
                            + \frac{1}{8} \left( d - 3 \right) \left( d - 5 \right)
                           \right) \,.
\label{K+}
\end{equation}
Furthermore one finds  that replacement of  $\mathfrak{so}(d-2)_L$ by $\mathfrak{so}(d-2)_J$ does not affect the generators $M_{\pm,0}$ and $\Delta$ in grade 0 subspace, $U_i$ and $U_i^\dag$ in grade $-1$ subspace, and $K_-$ in grade $-2$ subspace of $\mathfrak{so}(d,2)$. However it
leads to modifications of the grade $+1$ generators $W_i$ and $W_i^\dag$ involving $S_{ij}$, according to equation (\ref{SO(d,2)grade+1}):
\begin{equation}
\begin{split}
W_i
&= p \, a_i
  - \frac{i}{x}
    \left[
     \frac{1}{2} \left( d - 3 \right) a_i + i \left( L_{ij} + 2  \, S_{ij} \right) a_j
    \right]
\\
W_i^\dag
&= p \, a_i^\dag
   - \frac{i}{x}
     \left[
      \frac{1}{2} \left( d - 3 \right) a_i^\dag + i \left( L_{ij} + 2  \, S_{ij} \right) a_j^\dag
     \right]
\end{split}
\label{Grade+1Bosonic}
\end{equation}
Jacobi identities require that the spin generators $S_{ij}$ satisfy the constraint:
\begin{equation}
\Delta_{ij} = S_{ik} S_{jk} + S_{jk} S_{ik} - \frac{2}{(d-2)} \, \mathcal{S}^2 \, \delta_{ij} = 0
\label{SO(d-2)_Sconstraint}
\end{equation}
Remarkably, this constraint  (\ref{SO(d-2)_Sconstraint}) is precisely the condition that must be satisfied by the little group generators $S_{ij}$ of massless representations of the Poincar\'{e} group in $d$ dimensions that extend to the unitary irreducible representations of the conformal group $SO(d,2)$ \cite{Angelopoulos:1997ij,Laoues:1998ik}. That the minrep corresponds to a massless conformal scalar field and its deformations describe higher spin massless conformal fields in $d=4,5$ and $6$ dimensions were established in our previous work \cite{Fernando:2009fq,Fernando:2010dp,Fernando:2010ia,Fernando:2014pya}. The above analysis shows that this is true in all spacetime dimensions, namely there is a one-to-one correspondence between massless conformal fields in $d$-dimensional Minkowskian spacetimes and the minrep of $SO(d,2)$ and its deformations. It is interesting to note that each such deformation corresponds to a particular realization of the Calogero model or the conformal quantum mechanics with the distinguished $SU(1,1)_K$ subgroup acting as its symmetry group.

To summarize, of all  the generators  of $\mathfrak{so}(d,2)$:
\begin{equation}
\begin{split}
\mathfrak{so}(d,2)
&= \mathfrak{g}^{(-2)} \oplus \mathfrak{g}^{(-1)} \oplus
   \left[ \,
    \Delta \oplus \mathfrak{so}(d-2)_J \oplus \mathfrak{su}(1,1)_M \,
   \right]
   \oplus \mathfrak{g}^{(+1)} \oplus \mathfrak{g}^{(+2)}
\\
&= K_- \oplus \left( \, U_i \,,\, U_i^\dag \, \right) \oplus
   \left( \, \Delta \oplus J_{ij} \oplus M_{\pm,0} \, \right) \oplus
   \left( \, W_i \,,\, W_i^\dag \, \right) \oplus K_+
\end{split}
\end{equation}
only  $K_+$ and $W_i$ , $W_i^\dagger$ and $J_{ij}$ involve the spin operators $S_{ij}$.
We shall now give the quadratic Casimir of this deformed minrep of $\mathfrak{so}(d,2)$.  Noting that
\begin{equation}
\begin{split}
\mathcal{C}_2 \left[ \mathfrak{so}(d-2)_J \right]
&= J_{ij} J_{ij}
= \mathcal{J}^2
\\
\mathcal{C}_2 \left[ \mathfrak{su}(1,1)_M \right]
&= {M_0}^2 - \frac{1}{2} \left( M_+ M_- + M_- M_+ \right)
= \mathcal{M}^2
= \frac{1}{8} \, \mathcal{L}^2 + \frac{1}{16} \left( d - 2 \right) \left( d - 6 \right)
\\
\mathcal{C}_2 \left[ \mathfrak{su}(1,1)_K \right]
&= \Delta^2 - 2 \left( K_+ K_- + K_- K_+ \right)
= \mathcal{K}^2
\\
&= - \mathcal{J}^2 + \frac{1}{2} \, \mathcal{L}^2 + \frac{(d-6)}{(d-2)} \, \mathcal{S}^2 - \frac{1}{4} \left( d - 2 \right) \left( d - 6 \right)
\\
\left[ U W \right]
&= U_i W_i^\dag + W_i^\dag U_i - U_i^\dag W_i - W_i U_i^\dag
= 2 i \left( \mathcal{J}^2 - \mathcal{S}^2 \right)
  + i \left( d - 2 \right)^2
\end{split}
\end{equation}
 the quadratic Casimir of the deformed $\mathfrak{so}(d,2)$ takes the form
\begin{equation}
\begin{split}
\mathcal{C}_2 \left[ \mathfrak{so}(d,2) \right]
&= \mathcal{C}_2 \left[ \mathfrak{so}(d-2)_J \right]
   + 4 \, \mathcal{C}_2 \left[ \mathfrak{su}(1,1)_M \right]
   - \mathcal{C}_2 \left[ \mathfrak{su}(1,1)_K \right]
   + i \, \left[ UW \right]
\\
&= \frac{(d+2)}{(d-2)} \, \mathcal{S}^2
   - \frac{1}{2} \left( d^2 - 4 \right) \,.
\end{split}
\end{equation}
Note that it reduces to the quadratic Casimir of the undeformed minrep of $\mathfrak{so}(d,2)$ when $\mathcal{S}^2 = 0$.

In our previous work the generators of spin operators $S_{ij}$ were realized as bilinears of fermionic oscillators. For conformal groups in $d=3,4,5$ and $6$ dimensions this is natural since they admit extensions to simple superalgebras $OSp(N\,|\,4,\mathbb{R}), SU(2,2\,|\,N), F(4)$ and $OSp(8^*\,|\,2N)$, respectively. Deformations of the minrep can then be fitted  into unitary supermultiplets of the corresponding superalgebras. For $SU(2,2\,|\,N)$ and $OSp(8^*\,|\,2N)$ the corresponding supermultiplets turn out to be doubleton supermultiplets that were constructed and studied long time ago \cite{Gunaydin:1984fk,Gunaydin:1984wc}. Over the field of real and complex numbers there do not exist any simple conformal superalgebras beyond six dimensions that obey the usual spin and statistics connection and do not involve any extra tensorial generators \cite{Nahm:1977tg}. The construction of the representations of compact orthogonal groups over the Fock spaces of fermionic oscillators is reviewed in appendices \ref{app:S_ijForOdd} and \ref{app:S_ijForEven}. We shall discuss the application of this construction to the spin operators $S_{ij}$ and the study of the  relevant representations that satisfy (\ref{SO(d-2)_Sconstraint}) in odd and even dimensions separately.

%%%%%%%%%%%%%%%%%%%%%%%%%%%%%%%%%%%%%%%%%%%%%%%%%
%%%%% Deformations in odd dimensions %%%%%%%%%%%%%%%%%%%%%%%%%%%
%%%%%%%%%%%%%%%%%%%%%%%%%%%%%%%%%%%%%%%%%%%%%%%%%

\subsection{Deformation in odd dimensions}
\label{subsec:OddDeformations}

In our previous work on five-dimensional conformal algebra \cite{Fernando:2014pya}, we showed that $SO(5,2)$ admits only two singleton representations, which are the analogs of Dirac singletons in three dimensions, corresponding to five-dimensional massless scalar and spinorial conformal fields. The minrep of $SO(5,2)$ is the scalar singleton representation and the \emph{unique} ``deformation'' of the minrep, labelled by the spin 1/2 of the little group $SO(3)$, is the spinorial singleton representation. As discussed in appendix \ref{app:S_ijForOdd},  this property extends to all odd-dimensional conformal algebras $\mathfrak{so}(d,2)$ -- the minrep of $SO(d,2)$ for odd $d$ admits a single nontrivial  ``deformation''. Therefore we shall introduce a ``deformation parameter'' $\epsilon = 0$ or $1$, so that the ``undeformed'' minrep corresponds to $\epsilon = 0$, and the ``deformed" minrep corresponds to $\epsilon = 1$. As shown in appendix \ref{app:S_ijForOdd}, the generators $S_{ij}$ of $SO(d-2)_S$ describing this unique deformation can be realized in terms of a single set of fermionic oscillators transforming in the fundamental representation of $\mathfrak{u}((d-3)/2)$ subalgebra of $\mathfrak{so}(d-2)_S$ for odd $d$. More specifically the Lie algebra $\mathfrak{so}(d-2)_S$ has a 5-grading with respect to its subalgebra $\mathfrak{u}((d-3)/2)$:
\begin{equation}
\begin{split}
\mathfrak{so}(d-2)_S
&= \mathfrak{g}^{(-1)} \oplus \mathfrak{g}^{(-1/2)} \oplus \mathfrak{g}^{(0)} \oplus \mathfrak{g}^{(+1/2)} \oplus \mathfrak{g}^{(+1)}
\\
&= Z_{rs} \oplus Y_r \oplus T_{rs} \oplus Y_r^\dag \oplus Z_{rs}^\dag
\end{split}
\end{equation}
and its generators can be realized in terms of fermionic oscillators
as follows:
\begin{equation}
\begin{aligned}
T_{rs} &= \frac{1}{2} \left( \xi_r^\dag \xi_s - \xi_s \xi_r^\dag \right) \\
\end{aligned}
\qquad \qquad
\begin{aligned}
Y_r &= \frac{1}{\sqrt{2}} \xi_r \\
Y_r^\dag &= \frac{1}{\sqrt{2}} \xi_r^\dag
\end{aligned}
\qquad \qquad
\begin{aligned}
Z_{rs} &= \xi_r \xi_s \\
Z_{rs}^\dag &= - \xi_r^\dag \xi_s^\dag
\end{aligned}
\end{equation}
where $r,s,\dots =1,2,\dots (d-3)/2$.
Denoting the Fock vacuum as  $\ket{0}_F$, such that
\begin{equation}
\xi_r \ket{0}_F = 0
\qquad \qquad
(r=1,\dots,(d-3)/2)
\end{equation}
one finds that the entire fermionic Fock space spanned by the  states\footnote{$[r_1,r_2,\dots,r_n]$ denotes the complete anti-symmetrization with weight 1.}
\begin{equation}
\ket{0}_F \,\,,\,\,
\xi_{r_1}^\dag \ket{0}_F \,\,,\,\,
\xi_{[r_1}^\dag \xi_{r_2]}^\dag \ket{0}_F \,\,,\,\,
\xi_{[r_1}^\dag \xi_{r_2}^\dag \xi_{r_3]}^\dag \ket{0}_F \,\,,\,\,
\dots\dots\dots \,\,,\,\,
\xi_{[1}^\dag \dots \xi_{\frac{d-3}{2}]}^\dag \ket{0}_F
\end{equation}
form  the $2^{\left( \frac{d-3}{2} \right)}$-dimensional  spinor  irrep  of $SO(d-2)_S$.
The spin operators $S_{ij}$ can be recast  in terms of $2^n\times 2^n$ ($n=(d-3)/2)$ Euclidean Dirac gamma matrices $\gamma_i$ as
 \begin{equation}
S_{ij}= \frac{1}{4} [ \gamma_i, \gamma_j] \,.
 \end{equation}

In odd dimensions, the ``coupling constant''  $\mathcal{G}$ of the isotonic (singular) oscillator, which appears in the generator $K_+$, can be written in the form
\begin{equation}
\mathcal{G}
= \frac{1}{4} \, \mathcal{L}^2
   + \epsilon \, \mathcal{L} \cdot \mathcal{S}
   + \epsilon \, \frac{1}{2} \left( d - 3 \right)
   + \frac{1}{8} \left( d - 3 \right) \left( d - 5 \right)
\label{DeformedGodd}
\end{equation}
where we have used the value of the quadratic Casimir $\mathcal{S}^2$ from equation (\ref{S^2}).  The parameter $\epsilon =0$ for  the true minrep and $\epsilon = 1$ for the unique deformation.

Let us now give the K-type decomposition of the \emph{unique} ``deformation'' of the minrep of $SO(d,2)$ for all odd $d$.
Consider the tensor product space of the Fock spaces of the $a$-type bosonic oscillators, the $\xi$-type fermionic oscillators, and the state space of the singular oscillator. Let  $\ket{0}$ be the vacuum state annihilated by both $a_i$ and $\xi_r$:
\begin{equation}
a_i \ket{0} = 0
\qquad (i=1,\dots,d-2)
\qquad \qquad
\xi_r \ket{0} = 0
\qquad (r=1,\dots,(d-3)/2)
\end{equation}
With respect to the isotonic ``coupling constant'' $\mathcal{G}$ in equation (\ref{DeformedGodd}), this vacuum state has the eigenvalue
\begin{equation}
g \left( \epsilon \right) =  \epsilon \, \frac{1}{2} \left( d - 3 \right) + \frac{1}{8} \left( d - 3 \right) \left( d - 5 \right) \,.
\end{equation}
Therefore, the state of the form
\begin{equation}
\ket{\Omega_{(\epsilon)}} = x^{\alpha_{g(\epsilon)}} e^{-x^2/2} \ket{0}
\end{equation}
is annihilated by all  grade $-1$ operators in $\mathfrak{C}^-$ of $\mathfrak{so}(d,2)$, if $\alpha_{g(\epsilon)}$ satisfies:
\begin{equation}
\alpha_{g(\epsilon)} = \frac{1}{2} \left( 1 + \sqrt{\left( d - 4 \right)^2 + 4 \left( d - 3 \right) \epsilon} \right)
\end{equation}
More specifically we have
\begin{equation}
\alpha_{g(0)} = \frac{(d-3)}{2}
\qquad \qquad \qquad
\alpha_{g(1)}=\frac{(d-1)}{2} \,.
\end{equation}
The state $\ket{\Omega_{(0)}}$ is the lowest weight vector of the true minrep and is a singlet of $SO(d)$ which we denoted as
$|\psi_0^{\alpha_g}\rangle $  in section \ref{sec:SU(1,1)_K}.

By acting on the  state $\ket{\Omega_{(1)}}$ with all the generators of $SO(d)$, one obtains a set of states which we will denote collectively as $| \Psi^{\alpha_{g(1)}}_\sigma \rangle$, where $\sigma=1,\dots , 2^{\frac{(d-1)}{2}}$, transforming in the $2^{\left( \frac{d-1}{2} \right)}$-dimensional spinor representation of $SO(d)$, with the Dynkin labels $(0,0,\dots,0,1)$. They are eigenstates of the $AdS_{(d+1)}$ energy operator $H$ with the eigenvalue
\begin{equation}
E_{0}^{\alpha_{g(1)}} =  \frac{1}{2} \left( d - 1 \right) \,.
\end{equation}
Since the states $| \Psi^{\alpha_{g(1)}}_\sigma \rangle$ are annihilated by all the  operators in $\mathfrak{C}^-$ of $\mathfrak{so}(d,2)$, the deformed minrep of $SO(d,2)$ is also a unitary lowest weight representation, and the states $| \Psi^{\alpha_{g(1)}}_\sigma \rangle$ form the lowest K-type of the spinorial singleton.  All the other states of the ``particle basis'' of the spinorial singleton  are obtained from the set of states $| \Psi^{\alpha_{g(1)}}_\sigma \rangle$ by repeatedly acting with the deformed grade $+1$ operators in the $\mathfrak{C}^+$ subspace of $SO(d,2)$:
\begin{equation}
| \Psi^{\alpha_{g(1)}}_\sigma \rangle
\quad , \quad
\widetilde{B}_{M_1}^\dag \, | \Psi^{\alpha_{g(1)}}_\sigma \rangle
\quad , \quad
\widetilde{B}_{(M_1}^\dag \widetilde{B}_{M_2)_o}^\dag \, | \Psi^{\alpha_{g(1)}}_\sigma \rangle
\quad ,
\widetilde{B}_{(M_1}^\dag \widetilde{B}_{M_2}^\dag \widetilde{B}_{M_3)_o}^\dag \, | \Psi^{\alpha_{g(1)}}_\sigma \rangle
\quad , \quad
\dots\dots
\end{equation}
In Table \ref{Table:SpinorSingleton}, we give the K-type decomposition of  the unique deformed minrep (spinorial singleton) of $SO(d,2)$, for odd $d$.
The deformed minrep $(\epsilon =1)$ describes a massless spinor field in $d$ dimensions, and together with the true minrep $(\epsilon =0)$, which describes a massless scalar field, they exhaust the list of conformally massless fields in odd-dimensional spacetimes \cite{Angelopoulos:1997ij}.

%%%%%%%%%%%%%%%%%%%%%%%%%%%%%%%%%%%%%%%%%%%%%%%%
\begin{small}
\begin{longtable}[c]{|c|c|c|c|}
\kill
%%%%%%%%%%%%%%%%%%%%%%%%%%%%%%%%%%%%%%%%%%%%%%%%

\caption[The deformed minimal unitary representation (spinorial singleton) of $SO(d,2)$ for odd $d$]
{K-type decomposition of the deformed minimal unitary representation (spinorial singleton) of $SO(d,2)$, for odd $d$, with the lowest weight vector $| \Psi^{\alpha_{g(1)}}_\sigma \rangle$. The $AdS$ energy, and the dimension and the Dynkin labels of $SO(d)$ irrep at each level are given.
\label{Table:SpinorSingleton}} \\
\hline
 & & & \\
States & $AdS_{(d+1)}$ & Dim of & Dynkin \\
 & Energy & $SO(d)$ irrep & labels of \\
 & $E$ &  & $SO(d)$ irrep \\
 & & & \\
\hline
 & & & \\
\endfirsthead
%%%%%%%%%%%%%%%%%%%%%%%%%%%%%%%%%%%%%%%%%%%%%%%%
\caption[]{(continued)} \\
\hline
 & & & \\
States & $AdS_{(d+1)}$ & Dim of & Dynkin \\
 & Energy & $SO(d)$ irrep & labels of \\
 & $E$ &  & $SO(d)$ irrep \\
 & & & \\
\hline
\endhead
%%%%%%%%%%%%%%%%%%%%%%%%%%%%%%%%%%%%%%%%%%%%%%%%
 & & & \\
\hline
\endlastfoot
%%%%%%%%%%%%%%%%%%%%%%%%%%%%%%%%%%%%%%%%%%%%%%%%

$\ket{\Psi^{\alpha_{g(1)}}_\sigma}$ &
$\frac{1}{2} \left( d - 1 \right)$ &
$2^{\left( \frac{d-1}{2} \right)}$ &
$(0,0,\dots,0,1)$
\\[8pt]

\hline
 & & & \\

$\widetilde{B}_{M_1}^\dag \ket{\Psi^{\alpha_{g(1)}}_\sigma}$ &
$\frac{1}{2} \left( d + 1 \right)$ &
$2^{\left( \frac{d-1}{2} \right)} \cdot \frac{(d-1)!}{1!(d-2)!}$ &
$(1,0,\dots,0,1)$
\\[8pt]

\hline
 & & & \\

$\widetilde{B}_{(M_1}^\dag \widetilde{B}_{M_2)_o}^\dag \ket{\Psi^{\alpha_{g(1)}}_\sigma}$ &
$\frac{1}{2} \left( d + 3 \right)$ &
$2^{\left( \frac{d-1}{2} \right)} \cdot \frac{d!}{2!(d-2)!}$ &
$(2,0,\dots,0,1)$
\\[8pt]

\hline
 & & & \\

$\widetilde{B}_{(M_1}^\dag \widetilde{B}_{M_2}^\dag
   \widetilde{B}_{M_3)_o}^\dag \ket{\Psi^{\alpha_{g(1)}}_\sigma}$ &
$\frac{1}{2} \left( d + 5 \right)$ &
$2^{\left( \frac{d-1}{2} \right)} \cdot \frac{(d+1)!}{3!(d-2)!}$ &
$(3,0,\dots,0,1)$
\\[8pt]

\hline
 & & & \\

\vdots & \vdots & \vdots & \vdots \\ [8pt]

\hline
 & & & \\

$\widetilde{B}_{(M_1}^\dag \dots \widetilde{B}_{M_n)_o}^\dag
   \ket{\Psi^{\alpha_{g(1)}}_\sigma}$ &
$\frac{1}{2} \left( d + 2n - 1 \right)$ &
$2^{\left( \frac{d-1}{2} \right)} \cdot \frac{(d+n-2)!}{n!(d-2)!}$ &
$(n,0,\dots,0,1)$
\\[8pt]

\hline
 & & & \\

\vdots & \vdots & \vdots & \vdots \\[8pt]

%%%%%%%%%%%%%%%%%%%%%%%%%%%%%%%%%%%%%%%%%%%%%%%
\end{longtable}
\end{small}
%%%%%%%%%%%%%%%%%%%%%%%%%%%%%%%%%%%%%%%%%%%%%%%

%%%%%%%%%%%%%%%%%%%%%%%%%%%%%%%%%%%%%%%%%%%%%%%%%
%%%%%% Deformations in even dimensions %%%%%%%%%%%%%%%%%%%%%%%%%%
%%%%%%%%%%%%%%%%%%%%%%%%%%%%%%%%%%%%%%%%%%%%%%%%%

\subsection{Deformations in even dimensions}
\label{subsec:EvenDeformations}

As stated above, in four and six dimensions the minrep of the conformal group admits an infinite set of deformations which correspond to higher spin massless conformal fields \cite{Fernando:2009fq,Fernando:2010dp,Fernando:2010ia}. In four dimensions there exist a continuous infinity of massless conformal fields labelled by helicity. Similarly in six dimensions there exist a discrete infinity of massless conformal fields labelled by the spin of an $SU(2)$ subgroup of the little group $SO(4)$ of massless particles. Let us now show that the minrep of the even conformal group $SO(d,2)$ admits an infinite set of deformations for all $d>6$ as well.

In Appendix  \ref{app:S_ijForEven} we review the construction of the representations of $SO(2n)$ over the Fock spaces of an arbitrary number of fermionic oscillators transforming in the fundamental representation of the $\mathfrak{u}(n)$ subalgebra of $\mathfrak{so}(2n)$ and present their Gelfand-Zetlin as well as Dynkin labels following \cite{Gunaydin:1988kz}. We shall apply this construction to  the representations of the little group $SO(d-2)_S$ of massless particles, generated by $S_{ij}$. Using the 3-grading of the Lie algebra $\mathfrak{so}(d-2)$ with respect to its subalgebra $\mathfrak{u}((d-2)/2)$:
\begin{equation}
\begin{split}
\mathfrak{so}((d-2)/2)_S
&= \mathfrak{g}^{(-1)} \oplus \mathfrak{g}^{(0)} \oplus \mathfrak{g}^{(+1)}
\\
&= Z_{rs} \oplus T_{rs} \oplus Z_{rs}^\dag
\end{split}
\end{equation}
one realizes its generators as bilinears of fermionic oscillators as follows:
\begin{equation}
\begin{split}
Z_{rs}
&= \vec{\alpha}_r \cdot \vec{\beta}_s - \vec{\alpha}_s \cdot \vec{\beta}_r
   + \varepsilon \, \xi_r \xi_s
\\
T_{rs}
&= \vec{\alpha}_r^\dag \cdot \vec{\alpha}_s - \vec{\beta}_s \cdot \vec{\beta}_r^\dag
   + \frac{\varepsilon}{2} \left( \xi_r^\dag \xi_s - \xi_s \xi_r^\dag \right)
\\
Z_{rs}^\dag
&= - \vec{\alpha}_r^\dag \cdot \vec{\beta}_s^\dag + \vec{\alpha}_s^\dag \cdot \vec{\beta}_r^\dag
   - \varepsilon \, \xi_r^\dag \xi_s^\dag
\end{split}
\end{equation}
where $r,s,\dots =1,2,\dots ,(d-2)/2$ ; $\, \epsilon =0,1$  and $\vec{\alpha}_r \cdot \vec{\beta}_s = \sum_{K=1}^P \alpha_r(K) \beta_s(K) $ . The expressions of the generators $S_{ij}$ in terms of the above bilinears are given in appendix \ref{app:S_ijForEven} (equation (\ref{spinforeven})). The representations of even orthogonal groups $SO(d-2)$ that satisfy the constraint (\ref{SO(d-2)_Sconstraint}) have the Dynkin labels $(0,\dots,0,0,f)_D$ or $(0,\dots,0,f,0)_D$, where $f$ is a non-negative integer \cite{Angelopoulos:1997ij,Laoues:1998ik}. These irreducible representations have the  Gelfand-Zetlin labels
\begin{equation*}
\left( 0,\dots,0,0,f \right)_D \,=\, \left( \frac{f}{2} , \dots , \frac{f}{2} , \frac{f}{2} \right)_{GZ}
\end{equation*}
\begin{equation*}
\left( 0,\dots,0,f,0 \right)_D \,=\, \left( \frac{f}{2} , \dots , \frac{f}{2} , - \frac{f}{2} \right)_{GZ}
\end{equation*}
where $f=2P+\epsilon$ is the number of colors of fermionic oscillators transforming in the fundamental representation of the subgroup $U((d-2)/2)$.  For $d=4k+2$, where $k$ is a positive integer,
these massless irreps have the following lowest weight vectors in the fermionic Fock space:
\begin{equation}
\begin{split}
\left( 0,\dots,0,0,f \right)_D
&\quad \Longleftrightarrow \quad
\ket{0}
\\
\left( 0,\dots,0,f,0 \right)_D
&\quad \Longleftrightarrow \quad
\alpha_{(r_1}^\dag(1) \beta_{s_1}^\dag(1) \alpha_{r_2}^\dag(2) \beta_{s_2}^\dag(2) \dots \alpha_{r_P}^\dag(P) \beta_{s_P}^\dag(P) \xi_{t)}^\dag \ket{0}
\end{split}
\end{equation}
and $( r_1, s_1, \dots , t )$  denotes complete symmetrization of indices.\footnote{In the definition of the completely-symmetrized state, $\xi_t^\dag$ is absent if $\varepsilon=0$.} For $d=4k$, where $k$ is a positive integer, the roles of the lowest weight vectors above are reversed:
\begin{equation}
\begin{split}
\left( 0,\dots,0,0,f \right)_D
&\quad \Longleftrightarrow \quad
\alpha_{(r_1}^\dag(1) \beta_{s_1}^\dag(1) \alpha_{r_2}^\dag(2) \beta_{s_2}^\dag(2) \dots \alpha_{r_P}^\dag(P) \beta_{s_P}^\dag(P) \xi_{t)}^\dag \ket{0}
\\
\left( 0,\dots,0,f,0 \right)_D
&\quad \Longleftrightarrow \quad
\ket{0}
\end{split}
\end{equation}
Since $f = 2P + \varepsilon$, where $P=0,1,\dots$ and $\varepsilon=0,1$, we have an \emph{infinite} set of deformations of the minrep that are in one-to-one correspondence with the conformally massless representations of $SO(d,2)$ for even $d$, that remain irreducible under reduction to the Poincar\'e group in $d$ dimensions.

One finds that the quadratic Casimir $\mathcal{S}^2$ have the eigenvalues
\begin{equation}
S^2 = \frac{1}{4} (d-2) f (d+f-4)
\end{equation}
for the massless irreps of $SO(d-2)$ in even dimensions in agreement with \cite{Angelopoulos:1997ij,Laoues:1998ik}. Note that these eigenvalues depend only on $f$ and enter into
the ``coupling constant'' $\mathcal{G}$ that  appears in the generator $K_+$:
\begin{equation}
\mathcal{G}
= \frac{1}{4} \, \mathcal{L}^2
   + \mathcal{L} \cdot \mathcal{S}
   + \frac{2}{(d-2)} \mathcal{S}^2
   + \frac{1}{8} \left( d - 3 \right) \left( d - 5 \right)
\label{DeformedGeven}
\end{equation}

We shall now give the K-type decomposition of the deformations of the minrep of $SO(d,2)$ for all even $d$.
Let $\ket{0}_F$ be the vacuum of the Fock space of fermionic oscillators:
\begin{equation}
\alpha_r(K) \ket{0}_F = 0
\qquad \qquad
\beta_r(K) \ket{0}_F = 0
\qquad \qquad
\xi_r \ket{0}_F = 0
\end{equation}
where $r=1,\dots,n=1,\dots,\frac{d-2}{2}$ and $K=1,\dots,P$. By acting on this vacuum state with the corresponding fermionic creation operators $\alpha_r(K)^\dag$, $\beta_r(K)^\dag$, $\xi_r^\dag$, one can obtain a basis for this $2^{f \left( \frac{d-2}{2} \right)}$-dimensional Fock space. Now
consider the tensor product space of the Fock spaces of the $a$-type bosonic oscillators, $\alpha$-type, $\beta$-type and $\xi$-type fermionic oscillators, and the state space of the singular oscillator. Let  $\ket{0}$ be the tensor product of the vacuum state $\ket{0}_B$ annihilated by all $a_i$ and the fermionic Fock vacuum $\ket{0}_F$:
\begin{equation}
\begin{split}
&a_i \ket{0} = 0
\qquad \qquad \qquad \qquad \qquad (i=1,\dots,d-2)
\\
\alpha_r(K) \ket{0} = 0
\qquad
&\beta_r(K) \ket{0} = 0
\qquad
\xi_r \ket{0} = 0
\qquad (r=1,\dots,(d-2)/2)
\end{split}
\end{equation}
With respect to the isotonic ``coupling constant'' $\mathcal{G}$ in equation (\ref{DeformedGeven}), the vacuum state has the eigenvalue
\begin{equation}
g \left( f \right)
=  \frac{1}{2} f \left( d - 4 + f \right) + \frac{1}{8} \left( d - 3 \right) \left( d - 5 \right) \,.
\end{equation}
Therefore, the state
\begin{equation}
\ket{\Omega_{(f)}(0)} = x^{\alpha_{g(f)}} e^{-x^2/2} \ket{0}
\end{equation}
and the state
\begin{equation}
\ket{\Omega_{(f)}(S)} = x^{\alpha_{g(f)}} e^{-x^2/2} \alpha_{(r_1}^\dag(1) \beta_{s_1}^\dag(1) \alpha_{r_2}^\dag(2) \beta_{s_2}^\dag(2) \dots \alpha_{r_P}^\dag(P) \beta_{s_P}^\dag(P) \xi_{t)}^\dag \ket{0}
\end{equation}
are annihilated by all  grade $-1$ operators in $\mathfrak{C}^-$ of $\mathfrak{so}(d,2)$, if  $\alpha_{g(f)}$ is chosen to be
\begin{equation}
\alpha_{g(f)}
= \frac{1}{2} \left( d + 2 f - 3 \right) \,.
\end{equation}

Since
\begin{equation*}
\commute{\widetilde{B}_M}{\widetilde{M}_{NP}}
= i \left( \delta_{MN} \, \widetilde{B}_P - \delta_{MP} \, \widetilde{B}_N \right)
\end{equation*}
(see equation (\ref{SO(d,2)Compact})), all the states generated by the action of $\widetilde{M}_{NP}$ on $\ket{\Omega_{(f)}(0)}$ or $\ket{\Omega_{(f)}(S)}$ will be annihilated by $\widetilde{B}_M$. These states, generated by the action of $\widetilde{M}_{NP}$, form the lowest  K-type of a unitary lowest weight representation of $SO(d,2)$. These lowest K-types turn out to be the representations $\left( 0,\dots,0,0,f \right)_D$ or $\left( 0,\dots,0,f,0 \right)_D$ of $SO(d)$ for even $d$. They are all eigenstates of the $AdS$ energy operator $H$ that determines the compact 3-grading of $SO(d,2)$. These eigenvalues turn out to be equal for the two representations:
\begin{equation}
E_0(f)
= \frac{1}{2} \alpha_{g(f)} + \frac{1}{4} \left( d - 1 \right)
= \frac{1}{2} \left( d + f - 2 \right)
\end{equation}

In Tables \ref{Table:EvenDeformed0} and \ref{Table:EvenDeformedS}, we give the lowest K-type of all the unitary lowest weight representations of $SO(d,2)$ for even $d$ and their K-type decompositions.

%%%%%%%%%%%%%%%%%%%%%%%%%%%%%%%%%%%%%%%%%%%%%%%%
\begin{small}
\begin{longtable}[c]{|c|c|c|}
\kill
%%%%%%%%%%%%%%%%%%%%%%%%%%%%%%%%%%%%%%%%%%%%%%%%

\caption[Deformed minimal unitary representation obtained from $\ket{\Omega_{(f)}(0)}$ of $SO(d,2)$ for even $d$]
{K-type decomposition of the deformed minimal unitary representation of $SO(d,2)$, for even $d$, obtained from the lowest weight vector $\ket{\Omega_{(f)}(0)} = \, | \, E_0(f) \,,\, (0,\dots,0,0,f) \, \rangle$. The $AdS$ energy and the Dynkin labels of $SO(d)$ irrep at each level are given.
\label{Table:EvenDeformed0}} \\
\hline
 & & \\
States & $AdS_{(d+1)}$ & Dynkin \\
 & Energy & labels of \\
 & $E$ & $SO(d)$ irrep \\
 & & \\
\hline
 & & \\
\endfirsthead
%%%%%%%%%%%%%%%%%%%%%%%%%%%%%%%%%%%%%%%%%%%%%%%%
\caption[]{(continued)} \\
\hline
 & & \\
States & $AdS_{(d+1)}$ & Dynkin \\
 & Energy & labels of \\
 & $E$ & $SO(d)$ irrep \\
 & & \\
\hline
\endhead
%%%%%%%%%%%%%%%%%%%%%%%%%%%%%%%%%%%%%%%%%%%%%%%%
 & & \\
\hline
\endlastfoot
%%%%%%%%%%%%%%%%%%%%%%%%%%%%%%%%%%%%%%%%%%%%%%%%

$\ket{\Omega_{(f)}(0)}$ &
$\frac{1}{2} \left( d + f - 2 \right)$ &
% &
$(0,0,\dots,0,0,f)$
\\[8pt]

\hline
 & & \\

$\widetilde{B}_{M_1}^\dag \ket{\Omega_{(f)}(0)}$ &
$\frac{1}{2} \left( d + f \right)$ &
% &
$(1,0,\dots,0,0,f)$
\\[8pt]

\hline
 & & \\

$\widetilde{B}_{(M_1}^\dag \widetilde{B}_{M_2)_o}^\dag \ket{\Omega_{(f)}(0)}$ &
$\frac{1}{2} \left( d + f + 2 \right)$ &
% &
$(2,0,\dots,0,0,f)$
\\[8pt]

\hline
 & & \\

$\widetilde{B}_{(M_1}^\dag \widetilde{B}_{M_2}^\dag \widetilde{B}_{M_3)_o}^\dag
   \ket{\Omega_{(f)}(0)}$ &
$\frac{1}{2} \left( d + f + 4 \right)$ &
% &
$(3,0,\dots,0,0,f)$
\\[8pt]

\hline
 & & \\

\vdots & \vdots & \vdots \\ [8pt]

\hline
 & & \\

$\widetilde{B}_{(M_1}^\dag \dots \widetilde{B}_{M_n)_o}^\dag \ket{\Omega_{(f)}(0)}$ &
$\frac{1}{2} \left( d + f + 2n - 2 \right)$ &
% &
$(n,0,\dots,0,0,f)$
\\[8pt]

\hline
 & & \\

\vdots & \vdots & \vdots \\[8pt]

%%%%%%%%%%%%%%%%%%%%%%%%%%%%%%%%%%%%%%%%%%%%%%%
\end{longtable}
\end{small}
%%%%%%%%%%%%%%%%%%%%%%%%%%%%%%%%%%%%%%%%%%%%%%%

%%%%%%%%%%%%%%%%%%%%%%%%%%%%%%%%%%%%%%%%%%%%%%%%
\begin{small}
\begin{longtable}[c]{|c|c|c|c|}
\kill
%%%%%%%%%%%%%%%%%%%%%%%%%%%%%%%%%%%%%%%%%%%%%%%%

\caption[Deformed minimal unitary representation obtained from $\ket{\Omega_{(f)}(S)}$ of $SO(d,2)$ for even $d$]
{K-type decomposition of the deformed minimal unitary representation of $SO(d,2)$, for even $d$, obtained from the lowest weight vector $\ket{\Omega_{(f)}(S)} = \, | \, E_0(f) \,,\, (0,\dots,0,f,0) \, \rangle$. The $AdS$ energy and the Dynkin labels of $SO(d)$ irrep at each level are given.
\label{Table:EvenDeformedS}} \\
\hline
 & & \\
States & $AdS_{(d+1)}$ & Dynkin \\
 & Energy & labels of \\
 & $E$ & $SO(d)$ irrep \\
 & & \\
\hline
 & & \\
\endfirsthead
%%%%%%%%%%%%%%%%%%%%%%%%%%%%%%%%%%%%%%%%%%%%%%%%
\caption[]{(continued)} \\
\hline
 & & \\
States & $AdS_{(d+1)}$ & Dynkin \\
 & Energy & labels of \\
 & $E$ & $SO(d)$ irrep \\
 & & \\
\hline
\endhead
%%%%%%%%%%%%%%%%%%%%%%%%%%%%%%%%%%%%%%%%%%%%%%%%
 & & \\
\hline
\endlastfoot
%%%%%%%%%%%%%%%%%%%%%%%%%%%%%%%%%%%%%%%%%%%%%%%%

$\ket{\Omega_{(f)}(S)}$ &
$\frac{1}{2} \left( d + f - 2 \right)$ &
% &
$(0,0,\dots,0,f,0)$
\\[8pt]

\hline
 & & \\

$\widetilde{B}_{M_1}^\dag \ket{\Omega_{(f)}(S)}$ &
$\frac{1}{2} \left( d + f \right)$ &
% &
$(1,0,\dots,0,f,0)$
\\[8pt]

\hline
 & & \\

$\widetilde{B}_{(M_1}^\dag \widetilde{B}_{M_2)_o}^\dag \ket{\Omega_{(f)}(S)}$ &
$\frac{1}{2} \left( d + f + 2 \right)$ &
% &
$(2,0,\dots,0,f,0)$
\\[8pt]

\hline
 & & \\

$\widetilde{B}_{(M_1}^\dag \widetilde{B}_{M_2}^\dag \widetilde{B}_{M_3)_o}^\dag
   \ket{\Omega_{(f)}(S)}$ &
$\frac{1}{2} \left( d + f + 4 \right)$ &
% &
$(3,0,\dots,0,f,0)$
\\[8pt]

\hline
 & & \\

\vdots & \vdots & \vdots \\ [8pt]

\hline
 & & \\

$\widetilde{B}_{(M_1}^\dag \dots \widetilde{B}_{M_n)_o}^\dag \ket{\Omega_{(f)}(S)}$ &
$\frac{1}{2} \left( d + f + 2n - 2 \right)$ &
% &
$(n,0,\dots,0,f,0)$
\\[8pt]

\hline
 & & \\

\vdots & \vdots & \vdots \\[8pt]

%%%%%%%%%%%%%%%%%%%%%%%%%%%%%%%%%%%%%%%%%%%%%%%
\end{longtable}
\end{small}

%%%%%%%%%%%%%%%%%%%%%%%%%%%%%%%%%%%%%%%%%%%%%%%%%
%%%%%% AdS_{(d+1)}/CFT_d bosonic higher spin algebra %%%%%%%%%%%%%%%%%%
%%%%%%%%%%%%%%%%%%%%%%%%%%%%%%%%%%%%%%%%%%%%%%%%%

\section{$AdS_{(d+1)}/CFT_d$ bosonic higher spin algebra}
\label{sec:bosonicHS}

Before constructing the higher spin algebra in $d$ dimensions, it is useful to write down the $\mathfrak{so}(d,2)$ generators in the ``canonical basis'', that we denote as  $M_{AB}$ ($A,B=0,1,\dots,d+1$).  In terms of the generators in the compact 3-grading with respect to $\mathfrak{C}^0 = \mathfrak{so}(d) \oplus \mathfrak{so}(2)$, they are defined as
\begin{equation}
\begin{aligned}
M_{0M} &= \frac{1}{2} \left( \widetilde{B}_M^\dag + \widetilde{B}_M \right)
\\
M_{MN} &= \widetilde{M}_{MN}
\end{aligned}
\qquad \qquad \qquad
\begin{aligned}
M_{0,d+1} &= H
\\
M_{M,d+1} &= \frac{i}{2} \left( \widetilde{B}_M^\dag - \widetilde{B}_M \right)
\end{aligned}
\end{equation}
They satisfy the canonical commutation relations:\begin{equation}
\commute{M_{AB}}{M_{CD}}
= i \left( \eta_{BC} M_{AD} - \eta_{AC} M_{BD} - \eta_{BD} M_{AC} + \eta_{AD} M_{BC} \right)
\end{equation}
where $\eta_{AB} = \mathrm{diag} \left( -,+,\dots,+,- \right)$.

In terms of the generators in the non-compact three grading $\mathcal{D}$, $\mathcal{M}_{\mu\nu}$, $\mathcal{P}_\mu$, $\mathcal{K}_\mu$ ($\mu,\nu=0,\dots,d-1$), the generators $M_{AB}$ can be expressed as:
\begin{equation}
\begin{aligned}
M_{\mu\nu} &= \mathcal{M}_{\mu\nu}
\\
M_{\mu,d} &= \frac{1}{2} \left( \mathcal{P}_\mu - \mathcal{K}_\mu \right)
\end{aligned}
\qquad \qquad \qquad
\begin{aligned}
M_{\mu,d+1} &= \frac{1}{2} \left( \mathcal{P}_\mu + \mathcal{K}_\mu \right)
\\
M_{d,d+1} &= - \mathcal{D}
\end{aligned}
\end{equation}

The $AdS_{(d+1)}/CFT_d$ higher spin algebra of Fradkin-Vasiliev type simply corresponds to the universal enveloping algebra of $SO(d,2)$, quotiented by its Joseph ideal \cite{Gunaydin:1989um,Vasiliev:1999ba,Eastwood:2002su,eastwood2005uniqueness,Govil:2013uta,Govil:2014uwa,Fernando:2014pya}.
We recall that the universal enveloping algebra $\mathscr{U}(\mfg)$ of a Lie algebra $\mfg$ is defined as the quotient:
\be
\mathscr{U}(\mfg) = \mathscr{G}/\mathscr{L}
\ee
where $\mathscr{G}$ is the associative algebra freely generated by elements of $\mfg$, and $\mathscr{L}$ is the ideal of $\mathscr{G}$ generated by elements of form $\{g h-h g - [g,h]\},  (g,h \in \mfg)$ where $[g,h]$ denotes the commutator. The Joseph ideal of a Lie algebra $\mathfrak{g}$ is a two-sided ideal inside its universal enveloping algebra that annihilates its minrep.

We shall denote the corresponding higher spin algebra as $hs(d,2)$:
\begin{equation}
hs(d,2) = \frac{\mathscr{U}(d,2)}{\mathscr{J}(d,2)}
\end{equation}
where $\mathscr{U}(d,2)$ is the universal enveloping algebra of $\mathfrak{so}(d,2)$ and $\mathscr{J}(d,2)$ is its Joseph ideal.

An explicit formula for the generators of the Joseph ideal of $SO(d,2)$ was given by Eastwood in \cite{eastwood2005uniqueness}:
\begin{equation}
\begin{split}
J_{ABCD}
&= M_{AB} M_{CD}
   - M_{AB} \circledcirc M_{CD}
   - \frac{1}{2} \commute{M_{AB}}{M_{CD}}
   + \frac{(d-2)}{4d(d+1)} \langle M_{AB} , M_{CD} \rangle \, \mathbf{1}
\\
&= \frac{1}{2} M_{AB} \cdot M_{CD}
   - M_{AB} \circledcirc M_{CD}
   + \frac{(d-2)}{4d(d+1)} \langle M_{AB} , M_{CD} \rangle \,  \mathbf{1} \,.
\end{split}
\label{Joseph}
\end{equation}
In the above expression, the symbol $\cdot$ denotes the symmetric product of two generators of $\mathfrak{so}(d,2)$:
\begin{equation}
M_{AB} \cdot M_{CD}
= M_{AB} M_{CD} + M_{CD} M_{AB} \,,
\end{equation}
the symbol $\circledcirc$ denotes the Cartan product of two generators of $\mathfrak{so}(d,2)$ \cite{eastwood2005cartan}:
\begin{equation}
\begin{split}
M_{AB} \circledcirc M_{CD}
&= \frac{1}{3} M_{AB} M_{CD}
   + \frac{1}{3} M_{DC} M_{BA}
\\
& \quad
   + \frac{1}{6} M_{AC} M_{BD}
   - \frac{1}{6} M_{AD} M_{BC}
   + \frac{1}{6} M_{DB} M_{CA}
   - \frac{1}{6} M_{CB} M_{DA}
\\
& \quad
   - \frac{1}{2d}
     \left(
      \eta_{BD} M_{AE} M_C^{~E} - \eta_{AD} M_{BE} M_C^{~E}
      + \eta_{AC} M_{BE} M_D^{~E} - \eta_{BC} M_{AE} M_D^{~E}
     \right)
\\
& \quad
   - \frac{1}{2d}
     \left(
      \eta_{BD} M_{CE} M_A^{~E} - \eta_{AD} M_{CE} M_B^{~E}
      + \eta_{AC} M_{DE} M_B^{~E} - \eta_{BC} M_{DE} M_A^{~E}
     \right)
\\
& \quad
   + \frac{1}{d(d+1)} M_{EF} M^{EF}
     \left( \eta_{AC} \eta_{BD} - \eta_{BC} \eta_{AD} \right) \,,
\end{split}
\end{equation}
and $\langle M_{AB} , M_{CD} \rangle$ denotes the Killing form of $SO(d,2)$:
\begin{equation}
\langle M_{AB} , M_{CD} \rangle
= - \frac{2d}{(d^2-4)} \, M_{EF} M_{GH}
  \left( \eta^{EG} \eta^{FH} - \eta^{EH} \eta^{FG} \right)
  \left( \eta_{AC} \eta_{BD} - \eta_{AD} \eta_{BC} \right)
\end{equation}
where $\eta^{AB}$ is the $SO(d,2)$ invariant metric.

The decomposition of the enveloping algebra $\mathscr{U}(d,2)$ under the  adjoint action of $\mathfrak{so}(d,2)$  is equivalent to computing symmetric products of the adjoint representation of $\mathfrak{so}(d,2)$  whose Young tableau is
\begin{equation}
M_{AB} \sim \parbox{10pt}{\YoungAA} \,\,\,.
\end{equation}
The symmetric tensor product of the adjoint representation of $SO(d,2)$ decomposes as follows:
\begin{equation}
\begin{split}
\left( \,\,\, \parbox{10pt}{\YoungAA} \,\, \otimes \,\, \parbox{10pt}{\YoungAA} \,\,\, \right)_S
&= \,\,\, \parbox{20pt}{\YoungBB}
  \,\,\, \oplus \,\,\,
  \parbox{10pt}{\YoungAAAA}
  \,\,\, \oplus \,\,\,
  \parbox{20pt}{\YoungB}
  \,\,\, \oplus \,\,\,
  \bullet
\\
\left( \frac{(d+2)(d+1)}{2} \otimes \frac{(d+2)(d+1)}{2} \right)_S \quad
&= \quad \frac{(d+2)(d+3)(d+4)(d-1)}{12}
\\
& \qquad \quad
   \oplus \quad
   \frac{(d+2)!}{(d-2)!4!}
\\
& \qquad \quad
   \oplus \quad
   \left( \frac{(d+2)(d+3)}{2} - 1 \right)
\\
& \qquad \quad
   \oplus \quad
   1
\end{split}
\label{adjointsymm}
\end{equation}
where $\bullet$ represents the quadratic Casimir of $SO(d,2)$. As was pointed out by Vasiliev in \cite{Vasiliev:1999ba}, the higher spin gauge fields in $AdS_{(d+1)}$ are described by traceless two-row Young tableaux:
\begin{equation} %\label{nbox}
\underbrace{\begin{picture}(100,30)(0,0)
\put(0,5){\line(0,1){20}}
\put(00,5){\line(1,0){100}}
\put(00,15){\line(1,0){100}}
\put(00,25){\line(1,0){100}}
\put(10,5){\line(0,1){20}}
\put(20,5){\line(0,1){20}}
\put(30,5){\line(0,1){20}}
\put(40,5){\line(0,1){20}}
\put(70,5){\line(0,1){20}}
\put(80,5){\line(0,1){20}}
\put(90,5){\line(0,1){20}}
\put(100,5){\line(0,1){20}}
\put(55,10){\makebox(0,0){$\cdots$}}
\put(55,20){\makebox(0,0){$\cdots$}}
\end{picture}}_{\mbox{$n$ boxes}}
\end{equation}
Therefore one has to remove all the representations on the right hand side of equation (\ref{adjointsymm}) except for the window diagram. This is precisely what one achieves by quotienting the enveloping algebra by the Joseph ideal. After quotienting the enveloping algebra by the Joseph ideal, the generators of the infinite higher spin algebra decompose under the finite dimensional subalgebra $\mathfrak{so}(d,2)$ as
\begin{equation}
\parbox{10pt}{\YoungAA} \quad \oplus \quad  \parbox{20pt}{\YoungBB} \quad \oplus \quad ...\quad \oplus\quad % \label{nbox}
%\underbrace{
\begin{picture}(100,20)(0,12)
\put(0,5){\line(0,1){20}}
\put(00,5){\line(1,0){100}}
\put(00,15){\line(1,0){100}}
\put(00,25){\line(1,0){100}}
\put(10,5){\line(0,1){20}}
\put(20,5){\line(0,1){20}}
\put(30,5){\line(0,1){20}}
\put(40,5){\line(0,1){20}}
\put(70,5){\line(0,1){20}}
\put(80,5){\line(0,1){20}}
\put(90,5){\line(0,1){20}}
\put(100,5){\line(0,1){20}}
\put(55,10){\makebox(0,0){$\cdots$}}
\put(55,20){\makebox(0,0){$\cdots$}}
\end{picture} \quad \oplus \quad \dots
\end{equation}

For the minimal unitary realization of $SO(d,2)$ obtained via the quasiconformal methods, the generators $J_{ABCD}$ of the Joseph ideal vanish identically as  operators. This was shown to be the case for $d=3,4,5,6$ dimensions earlier \cite{Govil:2013uta,Govil:2014uwa,Fernando:2014pya}. We have verified that this is true in all dimensions. Therefore the enveloping algebra of the minimal unitary realization of $SO(d,2)$ obtained by quantizing its quasiconformal realization leads directly to the bosonic higher spin $AdS_{(d+1)}$/Conf$_d$ algebra for all $d$.

As we established above in section \ref{sec:deformedSO(d,2)}, the minimal unitary representation of $SO(d,2)$ admits deformations, which are in one-to-one correspondence with the unitary irreducible representations that describe conformally massless fields in $d$-dimensional Minkowskian spacetimes. As was done in $d=4,5,6$ dimensions, we shall define deformations of the bosonic higher spin algebras as enveloping algebras of the deformed minreps.
For the deformed minreps, the representation $\parbox{10pt}{\YoungAAAA}$ in the symmetric tensor product of the adjoint representation does not vanish, while the representation $\parbox{20pt}{\YoungB}$ still vanishes:
\begin{equation}
\eta^{CD} M_{AC} \cdot M_{DB} = 0
\end{equation}
Therefore the gauge fields of deformed higher spin algebras will not consist only of those  that correspond to traceless two-row Young tableaux. The Young tableaux of the representations of higher spin gauge fields defined by the deformed higher spin algebras will be given by the symmetric tensor products of the adjoint $\parbox{10pt}{\YoungAA}$ of $SO(d,2)$ subject to the constraint:
\begin{equation*}
\left( \,\, \parbox{10pt}{\YoungAA} \, \otimes \, \parbox{10pt}{\YoungAA} \,\, \right)_{Sym}
= \,\, \parbox{20pt}{\YoungBB} \, \oplus \, \parbox{10pt}{\YoungAAAA}
\end{equation*}

To understand how the Joseph ideal gets deformed when going from the minrep to a deformed minrep, one decomposes the $SO(d,2)$-covariant generators $J_{ABCD}$ of the Joseph ideal with respect to the $SO(d-1,1)$ (Lorentz) subgroup.

First we have the lightlike conditions for momentum and special conformal generators:
\begin{equation}
P^2 = \mathcal{P}^\mu \mathcal{P}_\mu = 0
\quad , \quad
K^2 = \mathcal{K}^\mu \mathcal{K}_\mu = 0
\end{equation}
which hold for both the minrep and its deformations.

The generators of $J_{ABCD}$ that are quadratic in the generators of $SO(d,2)$ and vanish for the minrep are as follows:
\begin{equation}
\begin{split}
d \, \mathcal{D} \cdot \mathcal{D} + \mathcal{M}^{\mu\nu} \cdot M_{\mu\nu} + \frac{(d-2)}{2} \, \mathcal{P}^\mu \cdot \mathcal{K}_\mu
&= 0
\\
\mathcal{P}^\mu \cdot \left( \mathcal{M}_{\mu\nu} + \eta_{\mu\nu} \, \mathcal{D} \right)
&= 0
\\
\mathcal{K}^\mu \cdot \left( \mathcal{M}_{\mu\nu} - \eta_{\mu\nu} \, \mathcal{D} \right)
&= 0
\\
\eta^{\mu\nu} \, \mathcal{M}_{\mu\rho} \cdot \mathcal{M}_{\nu\sigma} - \mathcal{P}_{(\rho} \cdot \mathcal{K}_{\sigma)} + \left( d - 2 \right) \eta_{\rho\sigma}
&= 0
\\
\mathcal{M}_{\mu\nu} \cdot \mathcal{M}_{\rho\sigma} + \mathcal{M}_{\mu\sigma} \cdot \mathcal{M}_{\nu\rho} + \mathcal{M}_{\mu\rho} \cdot \mathcal{M}_{\sigma\nu}
&= 0
\\
\mathcal{D} \cdot \mathcal{M}_{\mu\nu} + \mathcal{P}_{[\mu} \cdot \mathcal{K}_{\nu]}
&= 0
\\
\mathcal{M}_{[\mu\nu} \cdot \mathcal{P}_{\rho]}
&= 0
\\
\mathcal{M}_{[\mu\nu} \cdot \mathcal{K}_{\rho]}
&= 0
\end{split}
\end{equation}
In the above equations, symmetrizations (round brackets) and anti-symmetrizations (square brackets) are done with weight one. However, we should stress that the dot product of two operators $A$ and $B$ is defined as $A \cdot B = AB + BA$.

When the expressions for the generators of the deformed minrep of $SO(d,2)$, which involve fermionic oscillators, are substituted in the above identities only the first three identities are satisfied and the remaining five identities get modified by ``spin''-dependent terms involving fermionic oscillators. For example, the fourth identity becomes:
\begin{equation}
\eta^{\mu\nu} \mathcal{M}_{\mu\rho} \cdot \mathcal{M}_{\nu\sigma}
- \mathcal{P}_{(\rho} \cdot \mathcal{K}_{\sigma)}
- \left[ \frac{2}{(d-2)} \mathcal{S}^2 - \left( d - 2 \right) \right] \eta_{\rho\sigma}
= 0
\end{equation}
where $\mathcal{S}^2$ is the Casimir of $SO(d-2)_S$. There are similar changes in the last four identities.

%%%%%%%%%%%%%%%%%%%%%%%%%%%%%%%%%%%%%%%%%%%%%%%%%
%%%%%% Conclusions %%%%%%%%%%%%%%%%%%%%%%%%%%%%%%%%%%%%
%%%%%%%%%%%%%%%%%%%%%%%%%%%%%%%%%%%%%%%%%%%%%%%%%

\section{Conclusions}
\label{sec:conclusions}

Extending our earlier work we have shown that there is a one-to-one correspondence between the minimal unitary representation of $SO(d,2)$ and its deformations and conformally massless fields in $d$-dimensional Minkowskian spacetimes. The minrep describes a massless conformal scalar field, while the deformations describe massless conformal fields that transform nontrivially under the little group $SO(d-2)$  in $d$-dimensions. Joseph ideal generators vanish identically as operators for the minrep obtained via the quasiconformal methods. Similarly  for the deformed minreps certain deformations of the Joseph ideal generators involving the spin terms vanish as operators.

Again extending earlier work, we have shown how the enveloping algebras of the deformed minreps lead directly to deformations of the standard bosonic higher spin algebras in all dimensions. These deformed higher spin algebras are defined as the quotient of the universal enveloping algebras of $SO(d,2)$ by the respective deformed Joseph ideals. Consequently, the higher spin algebras are in one-to-one correspondence with the conformally massless unitary representations of $SO(d,2)$ in $d$ dimensional Minkowskian spacetimes.

In dimensions $d \le 6$, the minimal unitary representation of the conformal group $SO(d,2)$ and its deformations can be fitted  into the minimal unitary supermultiplet of a Lie superalgebra that extends the Lie algebra of $SO(d,2)$ and its deformed supermultiplets. For example, the minimal unitary supermultiplet of $PSU(2,2|4)$ is the $N=4$ Yang-Mills multiplet, and the minimal unitary supermultiplet of $OSp(8^*|4)$ is the $(2,0)$ tensor multiplet in six dimensions. The enveloping algebras of the minreps of the conformal  Lie superalgebras in $d\leq 6$ and  deformations thereof define the higher spin superalgebras and their deformations in the respective dimensions. Since no simple Lie superalgebras over the real or complex field, that extend the conformal algebra $\mathfrak{so}(d,2)$ by a compact $R$-symmetry group \cite{Nahm:1977tg}, exist beyond six dimensions, we do not expect the minrep and its deformations beyond six dimensions to fit into unitary representations of a simple Lie superalgebra that obeys the usual spin and statistics connection.

It was shown in \cite{Maldacena:2011jn} that the three dimensional conformal theories with a unique stress energy tensor and a conserved higher spin current that are dual to Vasiliev type higher spin theories in $AdS_4$ are free. As was pointed out in \cite{Govil:2013uta,Govil:2014uwa}, since the minrep and its deformations obtained via the quasiconformal approach are isomorphic to doubleton representations in $d=4$ and $d=6$ which are obtained by realizing the generators as bilinears of  covariant twistorial oscillators,  the conformal field theories that are dual to Vasiliev type theories must be free or interacting but integrable. Since there is a one-to-one correspondence between the massless conformal fields and higher spin algebras and their deformations in all dimensions we expect such a duality to hold for any $d$.     The  latter possibility of a duality with an interacting and integrable conformal field theory  is suggested by the fact that quasiconformal realizations of the minrep and its deformations are non-linear, involving cubic and quartic terms  for $d>3$.\footnote{Conformal group in $d=3$ is isomorphic to the symplectic group $Sp(4,\mathbb{R})$ for which the quartic invariant vanishes and the quasiconformal realization reduces to twistorial bilinear construction.}  A concrete result in support of this comes from the results of \cite{Govil:2012rh} where the quasiconformal construction of the minrep of $D(2,1;\alpha)$ and its deformations were shown to describe the spectra of $N=8$ supersymmetric interacting quantum mechanical models that were studied using harmonic superspace techniques in \cite{Fedoruk:2009xf}. Furthermore the  quasiconformal realizations of minreps and their deformations all have a distinguished $SL(2,\mathbb{R})$ subalgebra  which is of the form that comes up in Calogero models or conformal  quantum mechanical models which are integrable. This suggests the possibility  that an interacting but integrable conformal field theory that is dual to an higher spin theory may be a  dimensionally reduced CFT.

{\bf Acknowledgements:} 
The research of M.G. was supported in part by the US Department of Energy under DOE Grant No: DE-SC0010534.

%%%%%%%%%%%%%%%%%%%%%%%%%%%%%%%%%%%%%%%%%%%%%%%%%
%%%%%% Appendices %%%%%%%%%%%%%%%%%%%%%%%%%%%%%%%%%%%%%
%%%%%%%%%%%%%%%%%%%%%%%%%%%%%%%%%%%%%%%%%%%%%%%%%

\appendix

\numberwithin{equation}{section}

\section*{Appendices}

%%%%%%%%%%%%%%%%%%%%%%%%%%%%%%%%%%%%%%%%%%%%%%%%%
%%%%%% App: SO(d-2) for odd d %%%%%%%%%%%%%%%%%%%%%%%%%%%%%%%
%%%%%%%%%%%%%%%%%%%%%%%%%%%%%%%%%%%%%%%%%%%%%%%%%

\section{Realization of $SO(d-2)_S$ for odd $d$}
\label{app:S_ijForOdd}

 In this appendix we review  the construction given in \cite{Gunaydin:1987hb} of the irreps of the compact odd orthogonal groups $SO(2n+1)$ in the Fock space of fermionic oscillators. The Lie algebra $\mathfrak{so}(2n+1)$ can be given a 5-graded decomposition with respect to its $\mathfrak{u}(n)$ subalgebra as
\begin{equation}
\begin{split}
\mathfrak{so}(2n+1)_S
&= \mathfrak{g}^{(-1)} \oplus \mathfrak{g}^{(-1/2)} \oplus \mathfrak{g}^{(0)} \oplus \mathfrak{g}^{(+1/2)} \oplus \mathfrak{g}^{(+1)}
\\
&= Z_{rs} \oplus Y_r \oplus T_{rs} \oplus Y_r^\dag \oplus Z_{rs}^\dag
\end{split}
\end{equation}
such that the operators belonging to various-grade subspaces satisfy the commutation relations:
\begin{equation}
\begin{split}
\commute{T_{rs}}{T_{tu}}
&= \delta_{st} \, T_{ru} - \delta_{ur} \, T_{ts}
\qquad \qquad
\commute{Z_{rs}^\dag}{Z_{tu}}
= \delta_{rt} \, T_{su} - \delta_{st} \, T_{ru} - \delta_{ru} \, T_{st} + \delta_{su} \, T_{rt}
\\
\commute{Y_r^\dag}{Y_s}
&= T_{rs}
\qquad \qquad
\commute{Y_r}{Y_s}
= Z_{rs}
\qquad \qquad \qquad \qquad
\commute{Y_r^\dag}{Y_s^\dag}
= - Z_{rs}^\dag
\\
\commute{T_{rs}}{Y_t^\dag}
&= \delta_{st} \, Y_r^\dag
\qquad
\commute{T_{rs}}{Z_{tu}^\dag}
= \delta_{st} \, Z_{ru}^\dag - \delta_{su} \, Z_{rt}^\dag
\qquad
\commute{Z_{rs}}{Y_t^\dag}
= \delta_{st} \, Y_r - \delta_{rt} \, Y_s
\\
\commute{T_{rs}}{Y_t}
&= - \delta_{rt} \, Y_s
\qquad
\commute{T_{rs}}{Z_{tu}}
= - \delta_{rt} \, Z_{su} + \delta_{ru} \, Z_{st}
\qquad
\commute{Z_{rs}^\dag}{Y_t}
= \delta_{rt} \, Y_s^\dag - \delta_{st} \, Y_r^\dag
\end{split}
\end{equation}

The generators $S_{ij}$ of $\mathfrak{so}(2n+1) = \mathfrak{so}(d-2)_S$ in the canonical basis can be expressed in terms of the above operators as follows:
\begin{subequations}
\begin{equation}
\begin{split}
S_{r,d-2}
&= \frac{1}{2} \left( Y_r + Y_r^\dag \right) - \frac{i}{2} \left( Y_r - Y_r^\dag \right)
\\
S_{\frac{d-3}{2}+r,d-2}
&= \frac{1}{2} \left( Y_r + Y_r^\dag \right) + \frac{i}{2} \left( Y_r - Y_r^\dag \right)
\end{split}
\end{equation}
\begin{equation}
\begin{split}
S_{r,s}
&= i \commute{S_{r,d-2}}{S_{s,d-2}}
= \frac{1}{2} \left( Z_{rs} + Z_{rs}^\dag \right) + \frac{i}{2} \left( T_{rs} - T_{sr} \right)
\\
S_{r,\frac{d-3}{2}+s}
&= i \commute{S_{r,d-2}}{S_{\frac{d-3}{2}+s,d-2}}
= \frac{i}{2} \left( Z_{rs} - Z_{rs}^\dag \right) - \frac{1}{2} \left( T_{rs} + T_{sr} \right)
\\
S_{\frac{d-3}{2}+r,\frac{d-3}{2}+s}
&= i \commute{S_{\frac{d-3}{2}+r,d-2}}{S_{\frac{d-3}{2}+s,d-2}}
= - \frac{1}{2} \left( Z_{rs} + Z_{rs}^\dag \right) + \frac{i}{2} \left( T_{rs} - T_{sr} \right)
\end{split}
\end{equation}
\end{subequations}
where $r,s=1,\dots,n=1,\dots,\frac{d-3}{2}$.

They satisfy the commutation relations of the algebra $\mathfrak{so}(d-2)_S$:
\begin{equation}
\commute{S_{ij}}{S_{kl}}
= i \left( \delta_{jk} S_{il} - \delta_{ik} S_{jl} - \delta_{jl} S_{ik} + \delta_{il} S_{jk} \right)
\end{equation}
as well as the constraint given in equation (\ref{SO(d-2)_Sconstraint}).

The ``singletonic'' realization of $\mathfrak{so}(2n+1)$ is obtained by realizing its generators in terms of a single set of fermionic oscillators $\xi_r$ and $\xi_r^\dag$ ($r=1,\dots,n$) that transforms in the fundamental representation of $\mathfrak{u}(n)$ and satisfies
\begin{equation}
\anticommute{\xi_r}{\xi_s^\dag} = \delta_{rs}
\qquad \qquad
\anticommute{\xi_r}{\xi_s} = \anticommute{\xi_r^\dag}{\xi_s^\dag} = 0
\label{xi_commutators}
\end{equation}
and
\begin{equation}
\begin{aligned}
T_{rs} &= \frac{1}{2} \left( \xi_r^\dag \xi_s - \xi_s \xi_r^\dag \right) \\
\end{aligned}
\qquad \qquad
\begin{aligned}
Y_r &= \frac{1}{\sqrt{2}} \xi_r \\
Y_r^\dag &= \frac{1}{\sqrt{2}} \xi_r^\dag
\end{aligned}
\qquad \qquad
\begin{aligned}
Z_{rs} &= \xi_r \xi_s \\
Z_{rs}^\dag &= - \xi_r^\dag \xi_s^\dag \,.
\end{aligned}
\end{equation}
The entire Fock space of these fermionic oscillators forms the spinor representation of $SO(2n+1)$ of dimension $2^n$ and satisfies the condition (\ref{SO(d-2)_Sconstraint}). Furthermore, the quadratic Casimir reduces to a $c$-number
\begin{equation}
\mathcal{C}_2 \left[ \mathfrak{so}(d-2)_S \right]
= \mathcal{S}^2
= S_{ij} S_{ij}
= \frac{1}{4} \left( d - 2 \right) \left( d - 3 \right)
\label{S^2}
\end{equation}
for this irrep.

All the irreps of $SO(2n+1)$ can be constructed over the Fock space of an arbitrary number of fermionic oscillators $\alpha_r(K)$, $\beta_r(K)$, $\psi(K)$, $\xi_r$ and their hermitian conjugates $\alpha_r^\dag(K)$, $\beta_r^\dag(K)$, $\psi^\dag(K)$, $\xi_r^\dag$, where $K=1,\dots,P$ is a ``color'' index. The general form of the generators of $\mathfrak{so}(2n+1)$ in the 5-graded decomposition were given in \cite{Gunaydin:1987hb}:
\begin{equation}
\begin{split}
Z_{rs}
&= \vec{\alpha}_r \cdot \vec{\beta}_s - \vec{\alpha}_s \cdot \vec{\beta}_r
   + \varepsilon \, \xi_r \xi_s
\\
Y_r
&= \vec{\alpha}_r \cdot \vec{\psi}^\dag - \vec{\beta}_r \cdot \vec{\psi}
   + \frac{\varepsilon}{\sqrt{2}} \xi_r
\\
T_{rs}
&= \vec{\alpha}_r^\dag \cdot \vec{\alpha}_s - \vec{\beta}_s \cdot \vec{\beta}_r^\dag
   + \frac{\varepsilon}{2} \left( \xi_r^\dag \xi_s - \xi_s \xi_r^\dag \right)
\\
Y_r^\dag
&= \vec{\psi} \cdot \vec{\alpha}_r^\dag - \vec{\psi}^\dag \cdot \vec{\beta}_r^\dag
   + \frac{\varepsilon}{\sqrt{2}} \xi_r^\dag
\\
Z_{rs}^\dag
&= - \vec{\alpha}_r^\dag \cdot \vec{\beta}_s^\dag + \vec{\alpha}_s^\dag \cdot \vec{\beta}_r^\dag
   - \varepsilon \, \xi_r^\dag \xi_s^\dag
\end{split}
\end{equation}
where $\varepsilon = 0,1$ and the `dot product' denotes summation over all pairs of oscillators labelled by the ``color'' or ``family'' index $K$, \emph{e.g.} $\vec{\alpha}_r \cdot \vec{\beta}_s = \sum_{K=1}^P \alpha_r(K) \beta_s(K)$. Hence we have a realization of $SO(2n+1)$ in terms of $f = 2P + \varepsilon$ sets of fermionic creation and annihilation operators transforming in the fundamental and anti-fundamental representation of $U(n)$, respectively.  They satisfy the canonical anticommutation relations
\begin{equation}
\begin{split}
\anticommute{\alpha_r(K)}{\alpha_s^\dag(L)}
&= \anticommute{\beta_r(K)}{\beta_s^\dag(L)}
= \delta_{rs} \, \delta_{KL}
\qquad \qquad
\anticommute{\psi(K)}{\psi^\dag(L)} = \delta_{KL}
\\
\anticommute{\alpha_r(K)}{\alpha_s(L)}
&= \anticommute{\beta_r(K)}{\beta_s(L)}
= \anticommute{\alpha_r(K)}{\beta_s(L)}
= \anticommute{\alpha_r(K)}{\beta_s^\dag(L)}
= 0
\\
\anticommute{\alpha_r(K)}{\psi(L)}
&= \anticommute{\alpha_r(K)}{\psi^\dag(L)}
= \anticommute{\beta_r(K)}{\psi(L)}
= \anticommute{\beta_r(K)}{\psi^\dag(L)}
= 0
\end{split}
\end{equation}
in addition to those in equation (\ref{xi_commutators}).

In the Fock space of all fermionic oscillators, the irreps of $SO(2n+1) = SO(d-2)$ are uniquely determined by a set of states, that transforms irreducibly under the subgroup $U(n)$ and are annihilated by the grade $-1/2$ generators $Y_r$, which we shall refer to as the ``lowest weight vector'' by an abuse of terminology since they contain a true lowest weight vector.

As was proven in \cite{Laoues:1998ik} for odd orthogonal groups $SO(d-2)$ the only irreducible nontrivial  representation that satisfies the condition (\ref{SO(d-2)_Sconstraint})
\begin{equation*}
\Delta_{ij}
= S_{ik} S_{jk} + S_{jk} S_{ik} - \frac{2}{(2n+1)} \, \delta_{ij} \, \mathcal{S}^2 =0
\end{equation*}
is  the $2^n$-dimensional spinor representation with $2n+1=d-2$.
This is simply the irrep obtained by the singletonic realization of $SO(2n+1)$ in terms of a single set of oscillators, \emph{i.e.} $P=0$ and $\varepsilon =1$ with the Fock vacuum $|0\rangle$  as the lowest weight vector. The full Fock space of $n$ fermionic oscillators $\xi_r^\dagger$ form the $2^n$-dimensional irreducible spinor representation of $SO(2n+1)$. Equivalently  the action of the generators $S_{ij}$ on the entire Fock space can be represented in terms of $2^n \times 2^n$ gamma matrices $\gamma_i$, which satisfy the Clifford algebra
\begin{equation}
\anticommute{\gamma_i}{\gamma_j} = 2 \, \delta_{ij} \,,
\end{equation}
as
\begin{equation}
S_{ij} = \frac{1}{4} \commute{\gamma_i}{\gamma_j}
\end{equation}
which does satisfy the condition (\ref{SO(d-2)_Sconstraint}).

%%%%%%%%%%%%%%%%%%%%%%%%%%%%%%%%%%%%%%%%%%%%%%%%%
%%%%%% App: SO(d-2) for even d %%%%%%%%%%%%%%%%%%%%%%%%%%%%%%
%%%%%%%%%%%%%%%%%%%%%%%%%%%%%%%%%%%%%%%%%%%%%%%%%

\section{Realization of $SO(d-2)_S$ for even $d$}
\label{app:S_ijForEven}

Fermionic oscillator realization of even orthogonal Lie algebras, as studied in \cite{Gunaydin:1988kz},    can be obtained from the odd-dimensional orthogonal Lie algebras by truncating away the generators $Y_r$ and $Y_r^\dag$ belonging to the grade $\pm 1/2$ subspaces reviewed above in appendix \ref{app:S_ijForOdd}.
This leads naturally to the  3-grading of $\mathfrak{so}(d-2)_S$ for even $d$ \cite{Gunaydin:1988kz}:
\begin{equation}
\begin{split}
\mathfrak{so}(2n)
&= \mathfrak{g}^{(-1)} \oplus \mathfrak{g}^{(0)} \oplus \mathfrak{g}^{(+1)}
\\
&= Z_{rs} \oplus T_{rs} \oplus Z_{rs}^\dag
\end{split}
\end{equation}
where
\begin{equation}
\begin{split}
Z_{rs}
&= \vec{\alpha}_r \cdot \vec{\beta}_s - \vec{\alpha}_s \cdot \vec{\beta}_r
   + \varepsilon \, \xi_r \xi_s
\\
T_{rs}
&= \vec{\alpha}_r^\dag \cdot \vec{\alpha}_s - \vec{\beta}_s \cdot \vec{\beta}_r^\dag
   + \frac{\varepsilon}{2} \left( \xi_r^\dag \xi_s - \xi_s \xi_r^\dag \right)
\\
Z_{rs}^\dag
&= - \vec{\alpha}_r^\dag \cdot \vec{\beta}_s^\dag + \vec{\alpha}_s^\dag \cdot \vec{\beta}_r^\dag
   - \varepsilon \, \xi_r^\dag \xi_s^\dag
\end{split}
\end{equation}

The generators $S_{ij}$ of $\mathfrak{so}(2n) = \mathfrak{so}(d-2)_S$ in the canonical basis can be expressed in terms of the above operators as follows:
\begin{equation}
\begin{split}
S_{rs}
&= \frac{1}{2} \left( Z_{rs} + Z_{rs}^\dag \right) + \frac{i}{2} \left( T_{rs} - T_{sr} \right)
\\
S_{r,\frac{d-2}{2}+s}
&= \frac{i}{2} \left( Z_{rs} - Z_{rs}^\dag \right) - \frac{1}{2} \left( T_{rs} + T_{sr} \right)
\\
S_{\frac{d-2}{2}+r,\frac{d-2}{2}+s}
&= - \frac{1}{2} \left( Z_{rs} + Z_{rs}^\dag \right) + \frac{i}{2} \left( T_{rs} - T_{sr} \right)
\end{split}
\label{spinforeven}
\end{equation}
where $r,s=1,\dots,n=1,\dots,\frac{d-2}{2}$ .

They satisfy the commutation relations of the algebra $\mathfrak{so}(d-2)_S$:
\begin{equation}
\commute{S_{ij}}{S_{kl}}
= i \left( \delta_{jk} S_{il} - \delta_{ik} S_{jl} - \delta_{jl} S_{ik} + \delta_{il} S_{jk} \right)
\end{equation}
with the quadratic Casimir
\begin{equation}
\mathcal{C}_2 \left[ \mathfrak{so}(d-2)_S \right]
= \mathcal{S}^2
= S_{ij} S_{ij} \,.
\end{equation}

One can obtain all the irreps of $SO(2n)$ over the Fock space of these fermionic oscillators \cite{Gunaydin:1988kz}. The irreps can be uniquely labelled by the $U(n)$ Young tableaux of a set of states in the Fock space that transform irreducibly under $U(n)$ and are annihilated by  the grade $-1$ generators $Z_{rs}$. We shall refer to this set of states simply as lowest $U(n)$  representation. The relationship between the Young tableaux labels of the lowest $U(n)$  representations and the Gelfand-Zetlin labels as well as Dynkin labels of the corresponding irreps of $SO(2n)$ were given in \cite{Gunaydin:1988kz}, which we reproduce in Tables \ref{Table:SO(2n)irrepsEven} and \ref{Table:SO(2n)irrepsOdd}.

%%%%%%%%%%%%%%%%%%%%%%%%%%%%%%%%%%%%%%%%%%%%%%%%
\begin{small}
\begin{longtable}[c]{|l|l|}
\kill
%%%%%%%%%%%%%%%%%%%%%%%%%%%%%%%%%%%%%%%%%%%%%%%%

\caption[LWRs of $SO(2n)$ for even $n$]
{The $U(n)$ Young-tableau label of a lowest $U(n)$ representation and the Gelfand-Zetlin and Dynkin labels of the corresponding irrep of $SO(2n)$ for even $n$. The integer $f = 2P + \varepsilon$ ($P=1,2,\dots $ and $\varepsilon =0,1$) designates the number of  colors of the fermionic oscillators  transforming in the fundamental representation of the $U(n)$ subgroup.
\label{Table:SO(2n)irrepsEven}} \\
%%%%%%%%%%%%%%%%%%%%%%%%%%%%%%%%%%%%%%%%%%%%%%%%
\hline
 & \\
$U(n)$ Young-tableau & $ \left[ l_1 \,,\, l_2 \,,\, \dots \,,\, l_n \right]_{YT}$ \\
of the lowest $U(n)$ rep & \\ \hline
 & \\
Gelfand-Zetlin labels & $\left( \frac{f}{2} - l_n \,,\, \frac{f}{2} - l_{n-1} \,,\, \dots \,,\, \frac{f}{2} - l_1 \right)_{GZ}$ \\
of the $SO(2n)$ irrep & \\ \hline
 & \\
Dynkin labels & $\left( l_{n-1} - l_n \,,\, l_{n-2} - l_{n-1} \,,\, \dots \,,\, l_2 - l_3 \,,\, l_1 - l_2 \,,\, f - l_1 - l_2 \right)_D$ \\
of the $SO(2n)$ irrep & \\
 & \\
\hline
%%%%%%%%%%%%%%%%%%%%%%%%%%%%%%%%%%%%%%%%%%%%%%%
\end{longtable}
\end{small}
%%%%%%%%%%%%%%%%%%%%%%%%%%%%%%%%%%%%%%%%%%%%%%%

%%%%%%%%%%%%%%%%%%%%%%%%%%%%%%%%%%%%%%%%%%%%%%%%
\begin{small}
\begin{longtable}[c]{|l|l|}
\kill
%%%%%%%%%%%%%%%%%%%%%%%%%%%%%%%%%%%%%%%%%%%%%%%%

\caption[LWRs of $SO(2n)$ for odd $n$]
{The $U(n)$ Young-tableau label of a lowest $U(n)$ representation and the Gelfand-Zetlin and Dynkin labels of the corresponding irrep of $SO(2n)$ for odd $n$. The integer $f = 2P + \varepsilon$ ($P=1,2,\dots $ and $\varepsilon =0,1$) designates the number of  colors of the fermionic oscillators  transforming in the fundamental representation of the $U(n)$ subgroup.
\label{Table:SO(2n)irrepsOdd}} \\
%%%%%%%%%%%%%%%%%%%%%%%%%%%%%%%%%%%%%%%%%%%%%%%%
\hline
 & \\
$U(n)$ Young-tableau & $ \left[ l_1 \,,\, l_2 \,,\, \dots \,,\, l_n \right]_{YT}$ \\
of the lowest $U(n)$ rep & \\ \hline
 & \\
Gelfand-Zetlin labels & $\left( \frac{f}{2} - l_n \,,\, \frac{f}{2} - l_{n-1} \,,\, \dots \,,\, \frac{f}{2} - l_2 \,,\, l_1 - \frac{f}{2} \right)_{GZ}$ \\
of the $SO(2n)$ irrep & \\ \hline
 & \\
Dynkin labels & $\left( l_{n-1} - l_n \,,\, l_{n-2} - l_{n-1} \,,\, \dots \,,\, l_2 - l_3 \,,\, f - l_1 - l_2 \,,\, l_1 - l_2 \right)_D$ \\
of the $SO(2n)$ irrep & \\
 & \\
\hline
%%%%%%%%%%%%%%%%%%%%%%%%%%%%%%%%%%%%%%%%%%%%%%%
\end{longtable}
\end{small}
%%%%%%%%%%%%%%%%%%%%%%%%%%%%%%%%%%%%%%%%%%%%%%%

As was shown in \cite{Laoues:1998ik}, the representations of even orthogonal groups $SO(2n)$ that satisfy the constraint (\ref{SO(d-2)_Sconstraint}) have the Dynkin labels $(0,\dots,0,0,f)_D$ or $(0,\dots,0,f,0)_D$, where $f$ is a non-negative integer. The representations $(0,\dots,0,0,f)_D$ of $SO(2n)$, with Gelfand-Zetlin labels
\begin{equation*}
\left( 0,\dots,0,0,f \right)_D \,=\, \left( \frac{f}{2} , \dots , \frac{f}{2} , \frac{f}{2} \right)_{GZ}
\end{equation*}
have:
\begin{itemize}
\item for even $n=\frac{(d-2)}{2}$, the lowest $U(n)$ representation generated from the Fock vacuum $\ket{0}$
\item for odd $n=\frac{(d-2)}{2}$, the lowest $U(n)$ representation generated from a state of the form
\begin{equation*}
\alpha_{(r_1}^\dag(1) \beta_{s_1}^\dag(1) \alpha_{r_2}^\dag(2) \beta_{s_2}^\dag(2) \dots \alpha_{r_P}^\dag(P) \beta_{s_P}^\dag(P) \xi_{t)}^\dag \ket{0}
\end{equation*}
\end{itemize}
 On the other hand, the representations $(0,\dots,0,f,0)_D$ of $SO(2n)$, with Gelfand-Zetlin labels
\begin{equation*}
\left( 0,\dots,0,f,0 \right)_D \,=\, \left( \frac{f}{2} , \dots , \frac{f}{2} , - \frac{f}{2} \right)_{GZ}
\end{equation*}
have:
\begin{itemize}
\item for  even $n=\frac{(d-2)}{2}$, the lowest $U(n)$ representation generated from a state of the form
\begin{equation*}
\alpha_{(r_1}^\dag(1) \beta_{s_1}^\dag(1) \alpha_{r_2}^\dag(2) \beta_{s_2}^\dag(2) \dots \alpha_{r_P}^\dag(P) \beta_{s_P}^\dag(P) \xi_{t)}^\dag \ket{0}
\end{equation*}
\item for  odd $n=\frac{(d-2)}{2}$, the lowest $U(n)$ representation generated from the Fock vacuum $\ket{0}$
\end{itemize}
 Note that $f = 2P + \varepsilon$ is the number of sets of fermionic creation and annihilation operators, and $( r_1 s_1 \dots t )$ of fermionic indices denotes complete symmetrization.\footnote{In the definition of the completely-symmetrized state, $\xi_t^\dag$ is absent if $\varepsilon=0$.}

Since $f = 2P + \varepsilon$, where $P=0,1,\dots$ and $\varepsilon=0,1$, we have an \emph{infinite} set of deformations of the minrep that are in one-to-one correspondence with the conformally massless representations of $SO(d,2)$.

%%%%%%%%%%%%%%%%%%%%%%%%%%%%%%%%%%%%%%%%%%%%%%%%
%%%%% App: Indices Used %%%%%%%%%%%%%%%%%%%%%%%%%%%%%%%%%
%%%%%%%%%%%%%%%%%%%%%%%%%%%%%%%%%%%%%%%%%%%%%%%%

\section{Indices used in the paper}
\label{app:Indices}

Here we give a list of indices we used in this paper and their ranges:
\begin{equation}
\begin{split}
i,j,k,l &= 1,\dots,d-2
\qquad \qquad
\mbox{$SO(d-2)$ vector indices}
\\
a,b,c,d &= 1,2
\qquad \qquad \qquad \quad \;\,
\mbox{$Sp(2,\mathbb{R})$ spinor indices}
\\
\mu,\nu,\rho,\tau &= 0,\dots,d-1
\qquad \qquad
\mbox{$SO(d-1,1)$ vector indices}
\\
M,N,P,Q &= 1,\dots,d
\qquad \qquad \quad \;\;
\mbox{$SO(d)$ vector indices}
\\
A,B,C,D,E,F &= 0,\dots,d+1
\qquad \qquad
\mbox{$SO(d,2)$ vector indices}
\\
r,s,t,u &= 1,\dots,\frac{d-3}{2}
\qquad \qquad
\mbox{$SO(d-2)_S$ spinor indices for odd $d$}
\\
r,s,t,u &= 1,\dots,\frac{d-2}{2}
\qquad \qquad
\mbox{$SO(d-2)_S$ spinor indices for even $d$}
\end{split}
\end{equation}

%%%%%%%%%%%%%%%%%%%%%%%%%%%%%%%%%%%%%%%%%%%%%%%%%
%%%%%% Bibliography %%%%%%%%%%%%%%%%%%%%%%%%%%%%%%%%%%%%
%%%%%%%%%%%%%%%%%%%%%%%%%%%%%%%%%%%%%%%%%%%%%%%%%


\begin{thebibliography}{10}

\bibitem{Fernando:2009fq}
S.~Fernando and M.~Gunaydin, ``{Minimal unitary representation of $SU(2,2)$ and
  its deformations as massless conformal fields and their supersymmetric
  extensions},'' \href{http://dx.doi.org/10.1063/1.3447773}{{\em J.Math.Phys.}
  {\bfseries 51} (2010) 082301},
\href{http://arxiv.org/abs/0908.3624}{{\ttfamily arXiv:0908.3624 [hep-th]}}.
%%CITATION = ARXIV:0908.3624;%%.

\bibitem{Fernando:2010dp}
S.~Fernando and M.~Gunaydin, ``{Minimal unitary representation of $SO*(8) =
  SO(6,2)$ and its $SU(2)$ deformations as massless $6D$ conformal fields and their
  supersymmetric extensions},''
  \href{http://dx.doi.org/10.1016/j.nuclphysb.2010.07.001}{{\em Nucl.Phys.}
  {\bfseries B841} (2010) 339--387},
\href{http://arxiv.org/abs/1005.3580}{{\ttfamily arXiv:1005.3580 [hep-th]}}.
%%CITATION = ARXIV:1005.3580;%%.

\bibitem{Fernando:2010ia}
S.~Fernando and M.~Gunaydin, ``{$SU(2)$ deformations of the minimal unitary
  representation of $OSp(8^*|2N)$ as massless $6D$ conformal supermultiplets},''
  \href{http://dx.doi.org/10.1016/j.nuclphysb.2010.10.019}{{\em Nucl.Phys.}
  {\bfseries B843} (2011) 784--815},
\href{http://arxiv.org/abs/1008.0702}{{\ttfamily arXiv:1008.0702 [hep-th]}}.
%%CITATION = ARXIV:1008.0702;%%.

\bibitem{Fernando:2014pya}
S.~Fernando and M.~Gunaydin, ``{Minimal unitary representation of 5$d$
  superconformal algebra $F(4)$ and $AdS_6/CFT_5$ higher spin
  (super)-algebras},''
  \href{http://dx.doi.org/10.1016/j.nuclphysb.2014.11.015}{{\em Nucl. Phys.}
  {\bfseries B890} (2014) 570--605},
\href{http://arxiv.org/abs/1409.2185}{{\ttfamily arXiv:1409.2185 [hep-th]}}.
%%CITATION = ARXIV:1409.2185;%%.

\bibitem{Gunaydin:1984fk}
M.~Gunaydin and N.~Marcus, ``{The Spectrum of the $S^5$ Compactification of the
  Chiral $N=2, D=10$ Supergravity and the Unitary Supermultiplets of $U(2,
  2/4)$},''
{\em Class. Quant. Grav.} {\bfseries 2} (1985) L11.
%%CITATION = CQGRD,2,L11;%%.

\bibitem{Gunaydin:1998jc}
M.~Gunaydin, D.~Minic, and M.~Zagermann, ``{Novel supermultiplets of
  $SU(2,2|4)$ and the $AdS_5/CFT_4 $ duality},''
  \href{http://dx.doi.org/10.1016/S0550-3213(99)00007-3}{{\em Nucl. Phys.}
  {\bfseries B544} (1999) 737--758},
\href{http://arxiv.org/abs/hep-th/9810226}{{\ttfamily arXiv:hep-th/9810226}}.
%%CITATION = HEP-TH/9810226;%%.

\bibitem{Gunaydin:1998sw}
M.~Gunaydin, D.~Minic, and M.~Zagermann, ``{$4D$ doubleton conformal theories,
  CPT and II B string on $AdS(5) \times S(5)$},''
  \href{http://dx.doi.org/10.1016/S0550-3213(98)00543-4}{{\em Nucl. Phys.}
  {\bfseries B534} (1998) 96--120},
\href{http://arxiv.org/abs/hep-th/9806042}{{\ttfamily arXiv:hep-th/9806042}}.
%%CITATION = HEP-TH/9806042;%%.

\bibitem{Gunaydin:1984wc}
M.~Gunaydin, P.~van Nieuwenhuizen, and N.~P. Warner, ``{General Construction of
  the Unitary Representations of Anti-De Sitter Superalgebras and the Spectrum
  of the $S^4$ Compactification of Eleven-Dimensional Supergravity},''
\href{http://dx.doi.org/10.1016/0550-3213(85)90129-4}{{\em Nucl. Phys.}
  {\bfseries B255} (1985) 63}.
%%CITATION = NUPHA,B255,63;%%.

\bibitem{Gunaydin:1999ci}
M.~Gunaydin and S.~Takemae, ``{Unitary supermultiplets of $OSp(8*|4)$ and the
  $AdS(7)/CFT(6)$ duality},''
  \href{http://dx.doi.org/10.1016/j.nuclphysb.2004.07.022}{{\em Nucl.Phys.}
  {\bfseries B578} (2000) 405--448},
\href{http://arxiv.org/abs/hep-th/9910110}{{\ttfamily arXiv:hep-th/9910110
  [hep-th]}}.
%%CITATION = HEP-TH/9910110;%%.

\bibitem{Fernando:2001ak}
S.~Fernando, M.~Gunaydin, and S.~Takemae, ``{Supercoherent states of
  $OSp(8^*|2N)$, conformal superfields and the $AdS(7)/CFT(6)$ duality},''
  \href{http://dx.doi.org/10.1016/S0550-3213(02)00076-7}{{\em Nucl.Phys.}
  {\bfseries B628} (2002) 79--111},
\href{http://arxiv.org/abs/hep-th/0106161}{{\ttfamily arXiv:hep-th/0106161
  [hep-th]}}.
%%CITATION = HEP-TH/0106161;%%.

\bibitem{Maldacena:1997re}
J.~M. Maldacena, ``{The Large $N$ limit of superconformal field theories and
  supergravity},'' {\em Adv.Theor.Math.Phys.} {\bfseries 2} (1998) 231--252,
\href{http://arxiv.org/abs/hep-th/9711200}{{\ttfamily arXiv:hep-th/9711200
  [hep-th]}}.
%%CITATION = HEP-TH/9711200;%%.

\bibitem{Witten:1995zh}
E.~Witten, ``{Some comments on string dynamics},''
\href{http://arxiv.org/abs/hep-th/9507121}{{\ttfamily arXiv:hep-th/9507121
  [hep-th]}}.
%%CITATION = HEP-TH/9507121;%%.

\bibitem{Govil:2013uta}
K.~Govil and M.~Gunaydin, ``{Deformed Twistors and Higher Spin Conformal
  (Super-)Algebras in Four Dimensions},''
\href{http://arxiv.org/abs/1312.2907}{{\ttfamily arXiv:1312.2907 [hep-th]}}.
%%CITATION = ARXIV:1312.2907;%%.

\bibitem{Govil:2014uwa}
K.~Govil and M.~Gunaydin, ``{Deformed Twistors and Higher Spin Conformal
  (Super-)Algebras in Six Dimensions},''
  \href{http://dx.doi.org/10.1007/JHEP07(2014)004}{{\em JHEP} {\bfseries 1407}
  (2014) 004},
\href{http://arxiv.org/abs/1401.6930}{{\ttfamily arXiv:1401.6930 [hep-th]}}.
%%CITATION = ARXIV:1401.6930;%%.

\bibitem{Gunaydin:1989um}
M.~Gunaydin, ``{Singleton and doubleton supermultiplets of space-time
  supergroups and infinite spin superalgebras},''. Invited talk given at
  Trieste Conf. on Supermembranes and Physics in (2+1)-Dimensions, Trieste,
  Italy, Jul 17-21, 1989 ed. by M. J. Duff, et. al., World Scientific (1990),
  p. 442-456.

\bibitem{Fradkin:1986qy}
E.~Fradkin and M.~A. Vasiliev, ``{Cubic Interaction in Extended Theories of
  Massless Higher Spin Fields},''
\href{http://dx.doi.org/10.1016/0550-3213(87)90469-X}{{\em Nucl.Phys.}
  {\bfseries B291} (1987) 141}.
%%CITATION = NUPHA,B291,141;%%.

\bibitem{Konshtein:1988yg}
S.~Konshtein and M.~A. Vasiliev, ``{Massless representations and admissibility
  condition for higher spin superalgebras},''
\href{http://dx.doi.org/10.1016/0550-3213(89)90301-5}{{\em Nucl.Phys.}
  {\bfseries B312} (1989) 402}.
%%CITATION = NUPHA,B312,402;%%.

\bibitem{Vasiliev:1999ba}
M.~A. Vasiliev, ``{Higher spin gauge theories: Star product and AdS space},''
\href{http://arxiv.org/abs/hep-th/9910096}{{\ttfamily arXiv:hep-th/9910096
  [hep-th]}}.
%%CITATION = HEP-TH/9910096;%%.

\bibitem{Eastwood:2002su}
M.~G. Eastwood, ``{Higher symmetries of the Laplacian},''
  \href{http://dx.doi.org/10.4007/annals.2005.161.1645}{{\em Annals Math.}
  {\bfseries 161} (2005) 1645--1665},
\href{http://arxiv.org/abs/hep-th/0206233}{{\ttfamily arXiv:hep-th/0206233
  [hep-th]}}.
%%CITATION = HEP-TH/0206233;%%.
%%%%%%%%%%%%%%%%
\bibitem{Vasiliev:2004cm}
M.~Vasiliev, ``{Higher spin superalgebras in any dimension and their
	representations},''
\href{http://dx.doi.org/10.1088/1126-6708/2004/12/046}{{\em JHEP} {\bfseries
		0412} (2004) 046},
\href{http://arxiv.org/abs/hep-th/0404124}{{\ttfamily arXiv:hep-th/0404124
		[hep-th]}}.
%%CITATION = HEP-TH/0404124;%%.


%%%%%%%%%%%%%%%%%%%%%%%%%%%%%%%%%%%%%%
\bibitem{Gunaydin:2006vz}
M.~G{\"u}naydin and O.~Pavlyk, ``A unified approach to the minimal unitary
  realizations of noncompact groups and supergroups,'' {\em JHEP} {\bfseries
  09} (2006) 050,
\href{http://arxiv.org/abs/hep-th/0604077}{{\ttfamily hep-th/0604077}}.
%%CITATION = HEP-TH 0604077;%%.

\bibitem{Joseph:1974}
A.~Joseph, ``Minimal realizations and spectrum generating algebras,'' {\em
  Comm. Math. Phys.} {\bfseries 36} (1974) 325.

\bibitem{Gunaydin:2005zz}
M.~G{\"u}naydin and O.~Pavlyk, ``Generalized spacetimes defined by cubic forms
  and the minimal unitary realizations of their quasiconformal groups,'' {\em
  JHEP} {\bfseries 08} (2005) 101,
\href{http://arxiv.org/abs/hep-th/0506010}{{\ttfamily hep-th/0506010}}.
%%CITATION = HEP-TH 0506010;%%.

\bibitem{Gunaydin:2001bt}
M.~G{\"u}naydin, K.~Koepsell, and H.~Nicolai, ``{T}he minimal unitary
  representation of ${E_{8(8)}}$,'' {\em Adv. Theor. Math. Phys.} {\bfseries 5}
  (2002) 923--946,
\href{http://arxiv.org/abs/hep-th/0109005}{{\ttfamily hep-th/0109005}}.
%%CITATION = HEP-TH 0109005;%%.

\bibitem{Gunaydin:2004md}
M.~G{\"u}naydin and O.~Pavlyk, ``{M}inimal unitary realizations of exceptional
  {U}-duality groups and their subgroups as quasiconformal groups,'' {\em JHEP}
  {\bfseries 01} (2005) 019,
\href{http://arxiv.org/abs/hep-th/0409272}{{\ttfamily hep-th/0409272}}.
%%CITATION = HEP-TH 0409272;%%.

\bibitem{joseph1976minimal}
A.~Joseph, ``The minimal orbit in a simple lie algebra and its associated
  maximal ideal,'' {\em Ann. Sci. {\'E}cole Norm. Sup.(4)} {\bfseries 9} no.~1,
  (1976) 1--29.

\bibitem{deAlfaro:1976je}
V.~de~Alfaro, S.~Fubini, and G.~Furlan, ``Conformal invariance in quantum
  mechanics,''
{\em Nuovo Cim.} {\bfseries A34} (1976) 569.
%%CITATION = NUCIA,A34,569;%%.

\bibitem{Calogero:1969af}
F.~Calogero and C.~Marchioro, ``{Lower bounds to the ground-state energy of
  systems containing identical particles},''
\href{http://dx.doi.org/10.1063/1.1664877}{{\em J.Math.Phys.} {\bfseries 10}
  (1969) 562--569}.
%%CITATION = JMAPA,10,562;%%.

\bibitem{Calogero:1970nt}
F.~Calogero, ``{Solution of the one-dimensional N body problems with quadratic
  and/or inversely quadratic pair potentials},''
\href{http://dx.doi.org/10.1063/1.1665604}{{\em J.Math.Phys.} {\bfseries 12}
  (1971) 419--436}.
%%CITATION = JMAPA,12,419;%%.

\bibitem{Casahorran:1995vt}
J.~Casahorran, ``{On a novel supersymmetric connection between harmonic and
  isotonic oscillators},'' {\em Physica A} {\bfseries 217} (1995) 429--39.
  DFTUZ-94-28.

\bibitem{MR2451306}
J.~F. Cari{\~n}ena, A.~M. Perelomov, M.~F. Ra{\~n}ada, and M.~Santander, ``A
  quantum exactly solvable nonlinear oscillator related to the isotonic
  oscillator,'' \href{http://dx.doi.org/10.1088/1751-8113/41/8/085301}{{\em J.
  Phys. A} {\bfseries 41} no.~8, (2008) 085301, 10}.

\bibitem{MR858831}
A.~M.~Perelomov, ``{Generalized coherent states and their applications},''
  {\em Texts and Monographs in Physics}, 320p, Springer-Verlag, Berlin, 1986.

\bibitem{Angelopoulos:1997ij}
E.~Angelopoulos and M.~Laoues, ``{Masslessness in $n$-dimensions},''
  \href{http://dx.doi.org/10.1142/S0129055X98000082}{{\em Rev.Math.Phys.}
  {\bfseries 10} (1998) 271--300},
\href{http://arxiv.org/abs/hep-th/9806100}{{\ttfamily arXiv:hep-th/9806100
  [hep-th]}}.
%%CITATION = HEP-TH/9806100;%%.

\bibitem{Laoues:1998ik}
M.~Laoues, ``{Massless particles in arbitrary dimensions},''
  \href{http://dx.doi.org/10.1142/S0129055X98000355}{{\em Rev. Math. Phys.}
  {\bfseries 10} (1998) 1079--1109},
\href{http://arxiv.org/abs/hep-th/9806101}{{\ttfamily arXiv:hep-th/9806101
  [hep-th]}}.
%%CITATION = HEP-TH/9806101;%%.

\bibitem{Nahm:1977tg}
W.~Nahm, ``{Supersymmetries and their representations},''
\href{http://dx.doi.org/10.1016/0550-3213(78)90218-3}{{\em Nucl. Phys.}
  {\bfseries B135} (1978) 149}.
%%CITATION = NUPHA,B135,149;%%.

\bibitem{Gunaydin:1988kz}
M.~Gunaydin and S.~Hyun, ``{Unitary lowest weight representations of the
  noncompact supergroup $OSp(2n|2m,R)$},''
\href{http://dx.doi.org/10.1063/1.528120}{{\em J.Math.Phys.} {\bfseries 29}
  (1988) 2367}.
%%CITATION = JMAPA,29,2367;%%.

\bibitem{eastwood2005uniqueness}
M.~Eastwood, P.~Somberg, and V.~Soucek, ``The uniqueness of the joseph ideal
  for the classical groups,'' \href{http://arxiv.org/abs/0512296}{{\ttfamily
  arXiv:0512296 [math]}}.

\bibitem{eastwood2005cartan}
M.~Eastwood, ``The cartan product,'' {\em Bulletin of the Belgian Mathematical
  Society-Simon Stevin} {\bfseries 11} no.~5, (2005) 641--651.

\bibitem{Maldacena:2011jn}
J.~Maldacena and A.~Zhiboedov, ``{Constraining Conformal Field Theories with A
  Higher Spin Symmetry},''
  \href{http://dx.doi.org/10.1088/1751-8113/46/21/214011}{{\em J. Phys.}
  {\bfseries A46} (2013) 214011},
\href{http://arxiv.org/abs/1112.1016}{{\ttfamily arXiv:1112.1016 [hep-th]}}.
%%CITATION = ARXIV:1112.1016;%%.

\bibitem{Govil:2012rh}
K.~Govil and M.~Gunaydin, ``{Minimal unitary representation of $D(2,1:\lambda)$
  and its $SU(2)$ deformations and $d=1, N=4 $ superconformal models},''
  \href{http://dx.doi.org/10.1016/j.nuclphysb.2012.12.006}{{\em Nucl.Phys.}
  {\bfseries B869} (2013) 111--130},
\href{http://arxiv.org/abs/1209.0233}{{\ttfamily arXiv:1209.0233 [hep-th]}}.
%%CITATION = ARXIV:1209.0233;%%.

\bibitem{Fedoruk:2009xf}
S.~Fedoruk, E.~Ivanov, and O.~Lechtenfeld, ``{New $D(2,1,\alpha)$ Mechanics with
  Spin Variables},'' \href{http://dx.doi.org/10.1007/JHEP04(2010)129}{{\em
  JHEP} {\bfseries 1004} (2010) 129},
\href{http://arxiv.org/abs/0912.3508}{{\ttfamily arXiv:0912.3508 [hep-th]}}.
%%CITATION = ARXIV:0912.3508;%%.

\bibitem{Gunaydin:1987hb}
M.~Gunaydin, ``{Unitary highest weight representations of noncompact
  supergroups},''
\href{http://dx.doi.org/10.1063/1.527920}{{\em J.Math.Phys.} {\bfseries 29}
  (1988) 1275--1282}.
%%CITATION = JMAPA,29,1275;%%.

\end{thebibliography}
\end{document}